\documentclass[twocolumn, tighten]{aastex62}
\usepackage{apjfonts}
\usepackage{amsmath}
\usepackage{mathrsfs}
\usepackage{graphicx}
\usepackage{color}

\DeclareMathAlphabet{\mathbi}{OT1}{ptm}{bx}{it}
\SetMathAlphabet\mathbi{bold}{OT1}{ptm}{bx}{it}

\defcitealias{Du2014}{Paper~I}	
\defcitealias{Wang2014a}{Paper~II}	
\defcitealias{Hu2015}{Paper~III}	
\defcitealias{Du2015}{Paper~IV}	
\defcitealias{Du2016a}{Paper~V}
\defcitealias{Du2016b}{Paper~VI}
\defcitealias{Xiao2018}{Paper~VII}

\def\cb{}

\shorttitle{Structure of the BLR and Mass of the BH in Mrk 142}
\shortauthors{Y.-R. Li et al.}

\begin{document}
\title{Supermassive Black Holes with High Accretion Rates in Active Galactic Nuclei. VIII. Structure of the Broad-Line Region and 
Mass of the Central Black Hole in Mrk 142\footnote{The softeware developed in this work is available at \url{https://github.com/LiyrAstroph/BRAINS}.}}

\author[0000-0001-5841-9179]{Yan-Rong Li}
\author{Yu-Yang Songsheng}
\author{Jie Qiu}
\author[0000-0001-8492-6369]{Chen Hu}
\author[0000-0002-5830-3544]{Pu Du}
\affiliation{Key Laboratory for Particle Astrophysics, Institute of High 
Energy Physics, Chinese Academy of Sciences, 19B Yuquan Road, 
Beijing 100049, China}

\author[0000-0003-2112-171X]{Kai-Xing Lu}
\affiliation{Yunnan Observatories, Chinese Academy of Sciences, Kunming 650011, China}

\author{Ying-Ke Huang}
\affiliation{Key Laboratory for Particle Astrophysics, Institute of High 
Energy Physics, Chinese Academy of Sciences, 19B Yuquan Road, 
Beijing 100049, China}

\author{Jin-Ming Bai}
\affiliation{Yunnan Observatories, Chinese Academy of Sciences, Kunming 650011, China}

\author[0000-0002-2121-8960]{Wei-Hao Bian}
\affiliation{Physics Department, Nanjing Normal University, Nanjing 210097, China}

\author[0000-0002-7330-4756]{Ye-Fei Yuan}
\affiliation{Department of Astronomy, University of Science and Technology of China, Hefei 230026, China}

\author[0000-0001-6947-5846]{Luis~C.~Ho}
\affiliation{ Kavli Institute for Astronomy and Astrophysics, Peking University, 
Beijing 100871, China}
\affiliation{Department of Astronomy, School of Physics, Peking University, 
Beijing 100871, China}

\author[0000-0001-7617-4232]{Jian-Min Wang}
\affiliation{Key Laboratory for Particle Astrophysics, Institute of High 
Energy Physics, Chinese Academy of Sciences, 19B Yuquan Road, 
Beijing 100049, China}
\affiliation{National Astronomical Observatories of China, Chinese 
Academy of Sciences, 20A Datun Road, Beijing 100012, China}
\affiliation{School of Astronomy and Space Science, University of Chinese Academy of Sciences, 
19A Yuquan Road, Beijing 100049, China}

\collaboration{(SEAMBH Collaboration)}

\email{liyanrong@mail.ihep.ac.cn, wangjm@mail.ihep.ac.cn}

\begin{abstract}
This is the eighth in a series of papers reporting on a large reverberation mapping campaign 
to measure black hole (BH) mass in high accretion rate active galactic nuclei (AGNs). We employ 
the recently developed dynamical modeling approach for broad-line regions (BLRs) based on
the method of Pancoast et al. to analyze the reverberation mapping dataset of
Mrk 142 observed in the first monitoring season. In this approach,  continuum variations are reconstructed
using a damped random walk process, and BLR structure is delineated using a flexible disk-like geometry, in which
BLR clouds move around the central BH with Keplerian orbits or inflow/outflow motion. 
The approach also includes the possibilities of anisotropic emission of BLR clouds, non-linear response of the line 
emission to the continuum, and different long-term trends in the continuum and emission-line variations. 
We implement the approach in a Bayesian framework that is apt for parallel computation and 
use a Markov Chain Monte Carlo technique to recover the parameters and 
uncertainties for the modeling, including mass of the central BH. We apply three BLR models with different prescriptions of 
BLR clouds distributions and find that the best model for fitting the data of Mrk 142 is a two-zone 
BLR model, consistent with the theoretical BLR model surrounding slim accretion disks. 
{\cb The best model yields a BH mass of $\log (M_\bullet/M_\odot)=6.23_{-0.45}^{+0.26}$, resulting in a virial factor 
of $\log f=-0.36_{-0.54}^{+0.33}$ for the full width at half maximum 
of the H$\beta$ line measured from the mean spectrum. The virial factors for the other measures of the H$\beta$ line width
are also presented. }
\end{abstract}

\section{Introduction}
Broad emission lines with widths of several thousands of kilometers per second are a
hallmark feature of the spectra of active galactic nuclei (AGNs).
The basic photoionization theory and the long-known fact that
both the lines and continuum emissions of AGNs vary on quite short timescales, 
ranging from days to months (e.g., \citealt{Burbidge1967, Cromwell1970}),
spurred the proposal of the widely used technique ``reverberation mapping'' (RM) by 
\cite{Blandford1982}. The principle underlying RM is that the broad-line region (BLR) 
is photoionized by the central ionizing continuum stemmed from the accretion disk and 
reproduces broad emission lines. The temporal behaviors of emission lines are thereby
blurred echoes of continuum variations with light-crossing delays.
The responses from different parts of the BLR have different time delays
and Doppler shifting velocities due to the motion of BLR gas in the gravitational potential
well of the central black hole (BH). Thus, by appropriately analyzing
the variation properties of continuum and emission lines, one can place
constraints on the geometry and kinematics of the BLR as well as the BH mass.

Due to the insufficient cadences or signal-to-noise ratios of spectroscopic data,
the major goals of most early RM experiments have largely focused on measuring characteristic 
time lags between the variations of emission lines and continuum, 
which correspond to the light-travel distance from the continuum source to the line-emitting gas.
The measured time lags are thereby used to deduce the sizes of BLRs (e.g., \citealt{Peterson1993}). 
The recognition that emission-line lags and line widths
follow the ``virial relationship'' lends RM technique to be a promising method 
for measuring BH mass in AGNs (e.g., \citealt{Wandel1997, Ho1999, Peterson1999, Peterson2004}).
Specifically, based on the virial theorem, virial
BH mass measurements are generally derived by combining the emission-line lags ($\tau$) and 
line widths ($\Delta V$) using a simple recipe (e.g., \citealt{Peterson2004})
\begin{equation}
M_{\rm  vir} = f \frac{c\tau (\Delta V)^2}{G},
\label{eqn_mvir}
\end{equation}
where $c$ is the speed of light and $G$ is the gravitational constant.
To connect this virial mass with true BH mass, a virial factor $f$ has to be invoked
on account of our ignorance of the geometry and kinematic of BLRs. The virial factor
is practically calibrated by comparing RM AGNs with measured
bulge stellar velocity dispersion against the well established 
$M_\bullet-\sigma_\star$ relation of local quiescent galaxies (e.g., \citealt{Onken2004, Ho2014}).
Clearly, the virial factor determined in this way applies in a statistical
sense and is subject to the intrinsic scatter of the $M_\bullet-\sigma_\star$ relation, 
which is about 0.3 dex (e.g., \citealt{Kormendy2013}). 
It is unknown yet if the virial factor has a common value for various AGN populations. 
Given the complicated structures of BLRs inferred from the previous velocity-binned
RMs (e.g., \citealt{Bentz2010, Denney2010, Grier2013}; \citealt{Du2016b}, hereafter \citetalias{Du2016b}), 
the virial factor is most likely to vary
from object to object.
Recent recalibration of the virial factor by \cite{Ho2014} using the comprehensively revised  
$M_\bullet-\sigma_\star$ relation of \cite{Kormendy2013} indeed showed that the virial 
factor seems to depend on the bulge type (classical or pseudo) of the host galaxies.
Therefore, invoking of the virial factor in traditional RM approach actually 
impedes on its own  a further improvement on BH mass measurements (\citealt{Krolik2001}).

The way around this weakness of traditional RM approach is, similar to the
BH mass measurements in quiescent galaxies through stellar or gas dynamics, 
to develop feasible dynamical models for BLRs and analyze velocity-resolved 
RM data to determine the BH mass self-consistently without the need to invoke 
the virial factor. Such an approach dates back to the early work of \cite{Bottorff1997}
who applied the outflow model of \cite{Emmering1992} to the RM database of the well-monitored 
Seyfert galaxy NGC 5548, in an attempt to constrain the central BH
mass and the BLR dynamics. With the aid of the development of the mathematical description
of AGN variability (\citealt{Kelly2009}), \cite{Pancoast2011}
constructed a Bayesian framework with a flexible BLR dynamical model for RM data analysis. 
In this new approach, BH mass and other BLR parameters (e.g., inclination angle and 
opening angle) are fully determined by comparing the model predictions with the observed 
time series of continuum and broad emission lines. 
\cite{Pancoast2014a} subsequently reinforced their previous framework
by incorporating more complicated phenomenological treatments on the anisotropy of BLR emissions and 
inflows and outflows. Based on the model of 
\cite{Pancoast2011}, \cite{Li2013} also carried out an independent implementation
that, additionally, includes the non-linear response of emission lines to the ionizing continuum.
Although the dynamical modeling approach is still at its early stage, the application
to several RM AGNs shows its remarkable capability for understanding BLR dynamics and measuring BH
mass (\citealt{Brewer2011b, Pancoast2012, Pancoast2014b, Grier2017, Pancoast2018, Williams2018}).

In 2012, we began a large RM observational project using the Lijiang 2.4m telescope
at Yunnan Observatories, aiming at monitoring a 
sample of selected high-accretion-rate AGN (hereafter dubbed as super-Eddington accreting 
massive black hole; SEAMBH) candidates with good cadences for measuring reliably BH 
mass and studying BLR physics. In particular, based on slim accretion disk model that 
describes BH accretion at high accretion rates, \cite{Wang2014b} demonstrated that the geometrically
thick funnel of inner slim disk produce anisotropic radiation field, which divides the surrounding BLR
into two regions with distinct incident ionizing photon fluxes. Such BLR structures are 
different from these of sub-Eddington accreting BHs, which are powered by standard 
geometrically thin accretion disks (\citealt{Shakura1973}). By combining with the previous 
RM sample that mainly consists of sub-Eddington accreting AGNs (\citealt{Bentz2013}), this RM project provides
important complementary objects to generate a homogeneous sample with a broad range of accretion rates.

The project has continued uninterrupted for six years and 
is still ongoing. In the first monitoring season (2012-2013),
nine objects are finally verified to show statistically significant H$\beta$ 
lags through cross-correlation analysis. The resulting datasets and 
time-lag analysis between the optical continuum 
and H$\beta$ emission line and \ion{Fe}{2} emission have been reported in the papers of
this series \citet[\citetalias{Du2014}]{Du2014}, \citet[\citetalias{Wang2014a}]{Wang2014a}, 
\citet[\citetalias{Hu2015}]{Hu2015}, and \citet[Paper VII]{Xiao2018}.
The datasets and time-lag measurements from the other monitoring seasons have also been reported in 
the papers of this series \citet[Paper IV]{Du2015}, \citet[Paper V]{Du2016a}, and \citet[Paper IX]{Du2018}. 
We make use of the spectroscopic datasets for the nine objects in the first monitoring season
and employ our developed dynamical modeling for BLRs to study the structure and dynamics of the H$\beta$ BLRs 
and derive the BH mass for these nine objects. This paper reports the first application to Mrk~142, which 
has the highest data quality among the nine objects.

The paper is organized as follows. We briefly describe the properties of observation data for Mrk 142 
in Section~2. Section~3 describes the method for reconstruction of the continuum light curve and 
Section~4 presents the methodology for BLR dynamical modeling and the Bayesian framework for 
inferring the model parameters. Section~5 summarizes the results from the dynamical modeling including the obtained 
BH mass, and also compares our results with these from the cross-correlation 
analysis. The discussions and conclusions are given in Sections~6 and 7, respectively.

\begin{figure}
\centering
\includegraphics[width=0.48\textwidth]{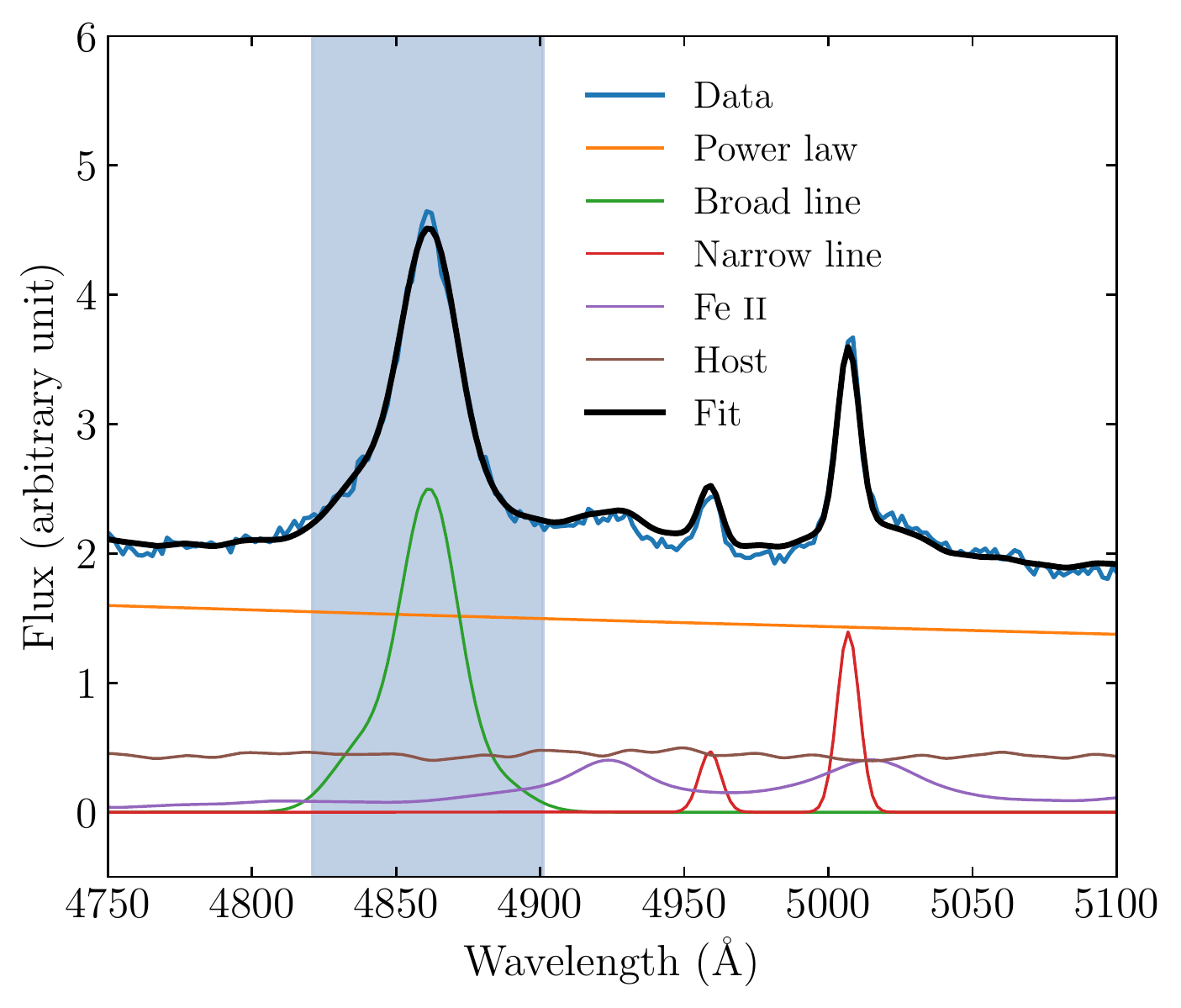}
\caption{\cb An example of spectral decomposition for Mrk 142. Shaded area indicates the wavelength range 
used in our BLR modeling.}
\label{fig_spec}
\end{figure}

\section{Observation Data}
Mrk 142 was 
spectroscopically and photometrically monitored between October 2012 and June 2013.
The details of the observations and data reduction and analysis 
were presented in \citetalias{Du2014} and \citetalias{Hu2015}. 
All the spectra were taken by simultaneously placing a nearby comparison star in the slit 
to achieve high-accuracy flux calibration. Since the host galaxies are extended and 
resolved, this observing strategy can, however, result in apparent flux variations 
of the host galaxy contamination due to variable seeing and miscentering.
A spectral decomposition scheme as described below helps to alleviate this effect 
(see Appendix A of \citetalias{Hu2015}).

To isolate the AGN continuum and H$\beta$ line,
a spectral decomposition was performed by
including a featureless AGN power-law continuum, \ion{Fe}{2} blends, various
emission lines (H$\beta$, [\ion{O}{3}], \ion{He}{2}, \ion{He}{1}, and several coronal lines),
and a host galaxy template (see \citetalias{Hu2015} for details). 
The \ion{Fe}{2} blends were fitted by the template from
\cite{Boroson1992} and the host galaxy template was chosen to be a single stellar
population model with an instantaneous burst of 11 Gyr and a metallicity of $Z=0.05$. The AGN continuum flux was measured as
the flux at 5100~{\AA} from the decomposed featureless power-law component. The H$\beta$ line profile
is obtained by subtracting the above-listed components from each spectrum.
The narrow H$\beta$ line component, which originates from the narrow-line region,
is not subtracted because of two reasons: first, Mrk 142 has fairly weak
narrow [\ion{O}{3}] lines, indicating that the narrow H$\beta$ lines are also weak;
second, the spectral resolution is about 500 $\rm km~s^{-1}$ in terms of the full width at half maximum 
(FWHM), too low to reliably decompose the narrow component (\citetalias{Hu2015}).
{\cb Figure~\ref{fig_spec} shows an example of spectral decomposition of an individual-night spectrum
for Mrk 142.}
\citetalias{Hu2015} also presented the H$\beta$ lag with respect to the 5100~{\AA} continuum through the 
cross-correlation method. Table~\ref{tab_sample} lists the overall 
properties of the observation data of Mrk 142.

%
\begin{deluxetable}{lcc}
\tabletypesize{\footnotesize}
\tablewidth{20cm}
\tablecolumns{3}
\tablecaption{Properties of the observation data of Mrk 142.\label{tab_sample}}
\tablehead{
\colhead{~~~~~~~~~~~~~~~~~~~} & \colhead{~~~~~~~~~~~~~~Value~~~~~~~~~~~~~~} & \colhead{~~~~~~~~~~~~~~Unit~~~~~~~~~~~~~~}
}
\startdata
(1) $z$         & 0.0449 & \nodata\\
(2) Dates  & 273-413 & day \\
(3) $N_{\rm con}$ &101 &  \nodata\\
(4) $N_{\rm H\beta}$ &101 & \nodata \\
(5) $F_{\rm var}$(5100\AA)        &   10.0\% & \nodata \\
(6) $F_{\rm var}$(H$\beta$)        &  6.8\%  & \nodata \\
(7) FWHM$_{\rm mean}$(H$\beta$)  & $1588\pm58$ & km s$^{-1}$\\
(8) $\sigma_{\rm mean}$(H$\beta$) & $972\pm12$ & km s$^{-1}$\\
(9) FWHM$_{\rm rms}$(H$\beta$)  & $1663\pm86$ & km s$^{-1}$\\
(10) $\sigma_{\rm rms}$(H$\beta$) & $1130\pm12$            & km s$^{-1}$\\
(11) $\Delta\lambda_{\rm dis}$ & $240\pm34$  & km s$^{-1}$ \\
(12) $\lambda$(H$\beta$)      & 4821-4901 &  \AA\\
(13) $\tau_{\rm cent}$  &  $7.9^{+1.2}_{-1.1}$  & day \\
(14) $\log(M_\bullet/M_\odot)$ & $6.59^{+0.07}_{-0.07}$ & \nodata 
\enddata
\tablecomments{The table rows are as follows: (1) redshift, (2) date range, JD-2,456,000, (3) number of epochs in the 5100~{\AA} continuum light curve,
(4) number of epochs of spectroscopy, (5) variability characteristic $F_{\rm var}$ of the 5100~{\AA} continuum, (6) variability characteristic 
$F_{\rm var}$ of the H$\beta$ fluxes, (7) H$\beta$ FWHM measured from the mean spectrum, {\cb (8) H$\beta$ line dispersion measured
from the mean spectrum, (9) H$\beta$ FWHM measured from the RMS spectrum, (10) H$\beta$ line dispersion measured from the RMS spectrum},
(11) spectral resolution given in terms of dispersion, (12) H$\beta$ wavelength range, 
set to be about three times the H$\beta$ FWHM, 
(13) H$\beta$ centroid time lag in the rest frame measured from CCF analysis by \citetalias{Hu2015},
and (14) BH mass, calculated using the centroid time lag and H$\beta$ FWHM measured from the mean spectrum 
with a virial factor $f_{\rm mean, FWHM}=1$.}
\end{deluxetable}

Three points merit emphasis regarding the datasets used for our dynamical modeling.
First, the spectra are aligned using [\ion{O}{3}]~$\lambda5007$ as the wavelength reference
to correct the night-to-night wavelength shifts.
Second, the wavelength range of the H$\beta$ line is adopted to be about three times the FWHM. 
Third, by comparing the observed spectrum of the comparison stars with stellar 
templates, \cite{Du2016b} determined the spectral resolution of the spectra at each individual epoch. 
We use the mean values for Mrk 142 as the input of the instrumental resolution to our dynamical 
modeling procedure (see below).

To correct for the redshift effect, we reduce the observation dates of the series
by a factor $(1+z)$, where $z=0.0449$ is the redshift of Mrk 142. Since the analysis 
only depends on the time differences between data points rather than the absolute times,
this manipulation equivalently converts the time series to the rest frame. 

\begin{figure*}[ht!]
\centering
\includegraphics[width=0.5\textwidth, angle=-90]{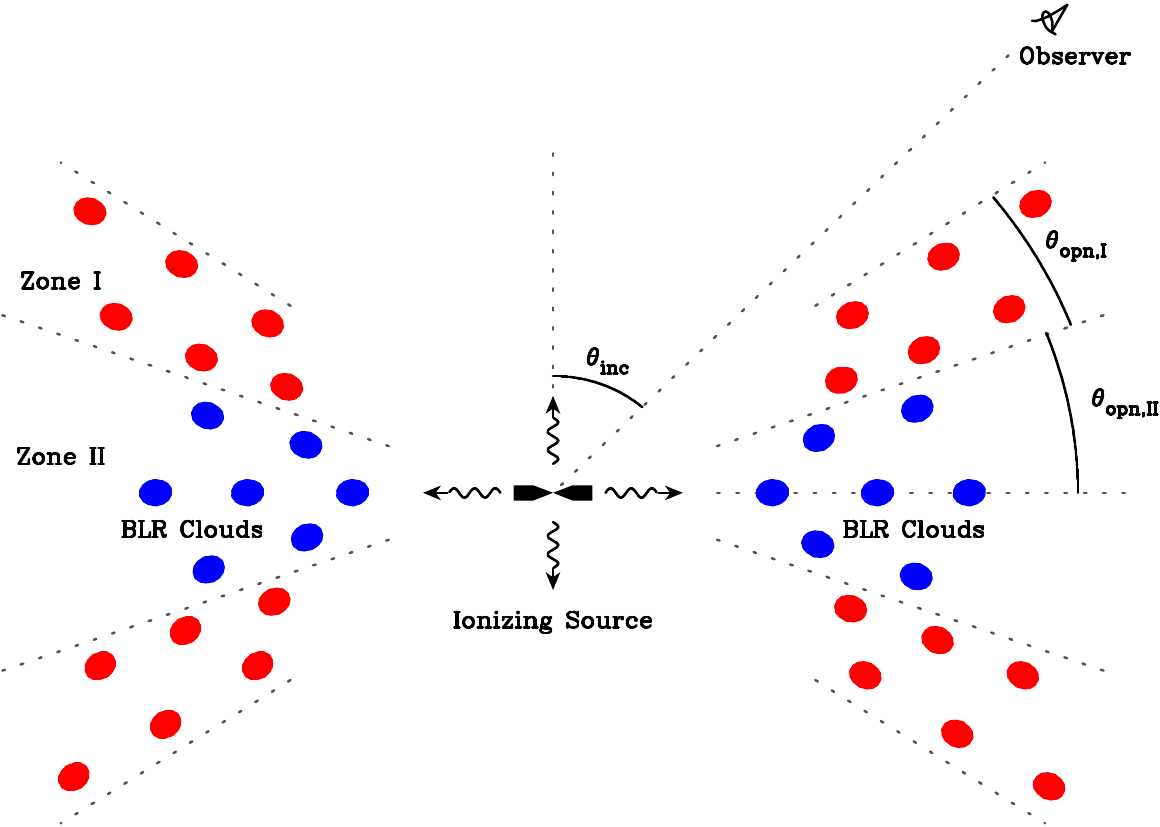}
\caption{Schematic of BLR model $M3$. $M3$ is a two-zone model
with the basic scenario that at high accretion rate regime, the geometrically thick funnel of 
the inner region of slim disks produces anisotropic ionizing radiation field that leads to two distinct BLR zones I and II.
We also construct two additional BLR models $M1$ and $M2$, which have the same disk-like geometry as zone II in model $M3$ 
(see Section 4 for details).}
\label{fig_sch}
\end{figure*}

%
\begin{deluxetable*}{cccccl}
\renewcommand{\arraystretch}{0.9}
\tabletypesize{\footnotesize }
\tablecolumns{6}
\tablewidth{1.0\textwidth}
\tablecaption{Parameters for models $M1$, $M2$, and $M3$.\label{tab_par}}
\tablehead{
  \colhead{Parameter}
& \colhead{Model\tablenotemark{a}}
& \colhead{~~~~~~~~Prior~~~~~~~~}
& \colhead{~~~~~~~~Range~~~~~~~~}
& \colhead{~~~~~~~~Unit~~~~~~~~}
& \colhead{Implication}
}
\startdata
\sidehead{Continuum}\hline
$\hat\sigma_{\rm d}$   &$M1, M2, M3$& Logarithmic       & $(10^{-3}, 10^{-1})$   & \nodata    & Long-term standard deviation of DRW variation\\
$\tau_{\rm d}$         &$M1, M2, M3$& Logarithmic       & $(1, 10^{4})$          & Day        & Typical timescale of DRW variation\\
$\mathbi{u_q}$         &$M1, M2, M3$& Gaussian          & \nodata                & \nodata    & Deviations of long-term trend of continuum fluxes \\
$\mathbi{u_s}$         &$M1, M2, M3$& Gaussian          & \nodata                & \nodata    & Deviations of continuum light curve\\\hline
\sidehead{BLR}\hline
$A$                &$M1, M2, M3$& Logarithmic   & (0.1, 10)       & \nodata    & Response coefficient of BLR\\
$\delta$           &$M1, M2, M3$& Uniform       & (-1, 3)         & \nodata    & Power-law index for non-linear response of BLR\\
$\mu$              &$M1, M3$& Logarithmic   & (0.1,100)    & ld   & Mean radius of BLR \\
$\beta$            &$M1, M3$& Uniform       & (0, 2)          & \nodata    & Shape of radial distribution of BLR particles \\
$F$                &$M1, M3$& Uniform       & (0, 1)          & \nodata    & Inner edge of BLR \\
$\mu_{\rm I}$      &$M3$& Logarithmic   & (0.1,100)    & ld   & Mean radius of BLR region I \\
$\beta_{\rm I}$     &$M3$& Uniform       & (0, 2)          & \nodata    & Shape of radial distribution of BLR particles for region I \\
$F_{\rm I}$         &$M3$& Uniform       & (0, 1)          & \nodata    & Inner edge of BLR region I\\
$\rho_{\rm I}$    & $M3$   & Uniform     & (0, 1)          &  \nodata   & Fraction of BLR particles in region I\\
$\alpha$           & $M2$    & Uniform       & (1, 3)          & \nodata    & Slope of double power law of radial distribution of BLR particles \\
$R_0$              & $M2$ & Logarithmic       & (0.1, 100)          & ld    & characteristic radius of double power law \\
$F_{\rm in}$       & $M2$        & Uniform       & (0, 1)          & \nodata    & Inner radius of double power law \\
$F_{\rm out}$      & $M2$         & Logarithmic       & (1, 10)          & \nodata    & Outer radius of double power law\\
$\theta_{\rm inc}$ &$M1, M2, M3$& Uniform       & (0, 90)         & Degree     & Inclination angle of BLR to the light of sight\\
$\theta_{\rm opn}$ &$M1, M2, M3$& Uniform       & (0, 90)         & Degree     & Opening angle of BLR\\
$\theta_{\rm opn, I}$ &$M3$& Uniform       & (0, 90)         & Degree     & Opening angle of BLR region I\\
$\kappa$           &$M1, M2, M3$& Uniform       & (-0.5, 0.5)     & \nodata    & Anisotropy of particle emission\\
$\gamma$           &$M1, M2, M3$& Uniform       & (1, 5)          & \nodata    & Clustering of BLR in the $\theta$ direction\\
$\xi$          &$M1, M2, M3$& Uniform       & (0, 1.5)        & \nodata    & Transparency of equatorial material\\
$M_\bullet$        &$M1, M2, M3$& Logarithmic   & $(10^5, 10^9)$   & $M_\odot$  & BH mass \\
$f_{\rm ellip}$    & $M1, M2, M3$  & Uniform       & (0, 1)          & \nodata    & Fraction of bound elliptical orbits\\
$f_{\rm flow}$     & $M1, M2, M3$ & Uniform       & (0, 1)          & \nodata    & Flag for determining inflowing or outflow orbits\\
$f_{\rm ellip, I}$    & $M3$  & Uniform       & (0, 1)          & \nodata    & Fraction of bound elliptical orbits for region I\\
$f_{\rm flow, I}$     & $M3$ & Uniform       & (0, 1)          & \nodata    & Flag for determining inflowing or outflow orbits for region I\\
$\sigma_{\rho, \rm circ}$     & $M1, M2, M3$ & Logarithmic       & (0.001, 0.1)          & \nodata    & Radial standard deviation around circular orbits\\
$\sigma_{\Theta, \rm circ}$   & $M1, M2, M3$   & Logarithmic       & (0.001, 1.0)          & \nodata    & Angular standard deviation around circular orbits\\
$\sigma_{\rho, \rm rad}$      & $M1, M2, M3$& Logarithmic       & (0.001, 0.1)          & \nodata    & Radial standard deviation around radial orbits\\
$\sigma_{\Theta, \rm rad}$    & $M1, M2, M3$  & Logarithmic       & (0.001, 1.0)          & \nodata    & Angular standard deviation around radial orbits\\
$\theta_{e}$      & $M1, M2, M3$& Uniform       & (0, 90)          & Degree    & Rotation angle of  of inflow or outflow orbits\\
$\sigma_{\rm turb}$  & $M1, M2, M3$    & Logarithmic       & (0.001, 0.1)          & \nodata    & Standard deviation of macroturbulent velocities\\
$\alpha_{\rm p}$   &$M1, M2, M3$& Uniform       & (-0.1, 0.1)     & \nodata    & Slope for different long-term trends in continuum and emission line\\
$\sigma_{{\rm s}, l}$  & $M1, M2, M3$     & \nodata   & $(1.0, 10.0)$     & \nodata    & Additive noise for emission line data\\
\enddata
\tablecomments{Parameters in bold are composed of an array. The unit of $\alpha_{\rm p}$ is (the unit of flux) $\times$ day$^{-1}$. 
The prior ranges of $A$ and $\alpha_{\rm p}$ are assigned in terms of 
the mean fluxes of the light curves normalized to unity. The prior for $\sigma_{{\rm s}, l}$ is set as $P(x) = 1/(1+x)$, 
where $x=\sigma_{{\rm s}, l}/\bar\sigma_{l}$ and $\bar\sigma_{l}$
is the mean measurement error of the emission line. $P(x)$ behaves like a uniform prior when $x\ll1$ and behaves like a
logarithmic prior when $x\gg1$ (\citealt{Gregory2011}).}
\tablenotetext{a}{This column indicates whether the parameters are included in the three models $M1$, $M2$, and $M3$.}
\end{deluxetable*}

\section{Continuum Modeling} 
\label{sec_con}
We use the damped random walk (DRW) model to describe the variability 
of the continuum fluxes (e.g., \citealt{Kelly2009, Zu2013} and references therein),
which allows us to interpolate and extrapolate continuum light curve in a statistical way
(\citealt{Pancoast2011}). In the DRW model, the covariance function between any two points 
at time $t_1$ and $t_2$ is given by
\begin{equation}
 S(t_1, t_2) = \sigma_{\rm d}^2 \exp\left(-\frac{|t_1-t_2|}{\tau_{\rm d}}\right),
 \label{eqn_cov}
\end{equation}
where $\sigma_{\rm d}$ is the long-term standard deviation of the variation and $\tau_{\rm d}$
is the typical timescale of variation. In this prescription, the variation of light curves 
at short time scale ($t\ll\tau_{\rm d}$) is $\sigma\sqrt{t/2\tau_{\rm d}}$. To relax the correlation between $\sigma_{\rm d}$
and $\tau_{\rm d}$, a new parameterization of 
$\hat\sigma_{\rm d}= \sigma_{\rm d}/\sqrt{\tau_{\rm d}}$ is used to replace $\sigma_{\rm d}$.

Let $\mathbi{s}$ denote the real underlying signal of the continuum variations to be inferred from 
observations. A set of measurements $\mathbi{y}_{\rm c}$ for a continuum light curve in a monitoring campaign
can be written as
\begin{equation}
\mathbi{y}_{\rm c}=\mathbi{s} + \mathbi{Lq} + \mathbi{n}_{\rm c}, 
\end{equation}
where $\mathbi{n}_{\rm c}$ represents the measurement noises and the term $\mathbi{Lq}$ represents 
a linearly varying trend in the light curve. 
Here $\mathbi{L}$ is a matrix of known coefficients and 
$\mathbi{q}$ is a vector of unknown linear coefficients (see \citealt{Rybicki1992} for details).
Assuming that the measurement noise $\mathbi{n}_{\rm c}$ is Gaussian and uncorrelated, 
the likelihood probability for $\mathbi{y}_{\rm c}$ is (\citealt{Rybicki1992, Li2013})
\begin{eqnarray}
P(\mathbi{y}_{\rm c}|\sigma_{\rm d}, \tau_{\rm d}, \mathbi{q}) &=& \frac{1}{\sqrt{(2\pi)^m |\mathbi{C}|}}
\exp\left\lgroup-\frac{(\mathbi{y}_{\rm c}-\mathbi{Lq})^{T}\mathbi{C}^{-1}(\mathbi{y}_{\rm c}-\mathbi{Lq})}{2}\right\rgroup,\nonumber\\
~~~
\end{eqnarray}
where superscript ``$T$'' denotes the transposition, $m$ is the number of data points, $\mathbi{C=S+N}$, $\mathbi{S}$ 
is the covariance matrix of $\mathbi{s}$ given by Equation~(\ref{eqn_cov}), and $\mathbi{N}$ is the covariance 
matrix of $\mathbi{n}_{\rm c}$. In our calculations, we by default include the zero-order linear trend. 
In this case, $\mathbi{L}$ is a vector with all unity elements and $\mathbi{q}$ is the long-term mean value of 
the light curve. This helps to remove the bias in modeling the light curve 
at epochs far from any data points (see discussion in \citealt{Rybicki1992}).

Given parameters ($\sigma_{\rm d}, \tau_{\rm d}, \mathbi{q}$), the probability of a signal $\mathbi{s}$ 
underlying a set of measurements $\mathbi{y}_{\rm c}$ is (e.g., \citealt{Rybicki1992, Zu2011})
\begin{eqnarray}
P(\mathbi{s}|\mathbi{y}_{\rm c})&\propto&\exp\left\lgroup- \frac{(\mathbi{s}-\mathbi{\hat s})^T\mathbi{Q}^{-1}(\mathbi{s}-\mathbi{\hat s})}{2}
-\frac{(\mathbi{q}-\mathbi{\hat q})^T\mathbi{C_q}^{-1}(\mathbi{q}-\mathbi{\hat q})}{2}\right.\nonumber\\
&&~~~~~~~~~\left. -\frac{(\mathbi{y}_{\rm c}-\mathbi{L\hat q})^T\mathbi{C}^{-1}(\mathbi{y}_{\rm c}-\mathbi{L\hat q})}{2}\right\rgroup,
\label{eqn_signal}
\end{eqnarray}
where 
\begin{eqnarray}
&&\mathbi{\hat s} = \mathbi{SC}^{-1}(\mathbi{y}_{\rm c}-\mathbi{Lq}),\label{eqn_s}\\
&&\mathbi{\hat q} = \mathbi{C_q}\mathbi{L}^T\mathbi{C}^{-1}\mathbi{y}_{\rm c},\label{eqn_q}\\
&&\mathbi{C_q} = (\mathbi{L}^T\mathbi{C}^{-1}\mathbi{L})^{-1},\\
&&\mathbi{Q} = [\mathbi{S}^{-1}+\mathbi{N}^{-1}]^{-1}.\label{eqn_Q}
\end{eqnarray}
Equation~(\ref{eqn_signal}) indicates that the signal $\mathbi{s}$ is a Gaussian process with mean $\mathbi{\hat s}$
and covariance matrix $\mathbi{Q}$ (\citealt{Rybicki1992}). One can thereby generate a signal $\mathbi{s}$ by adding to $\mathbi{\hat s}$
a Gaussian process with zero mean and covariance matrix $\mathbi{Q}$ given by Equation~(\ref{eqn_Q}).
Similarly, the probability of $\mathbi{q}$ is a Gaussian with mean $\mathbi{\hat q}$ and covariance matrix $\mathbi{C}_{q}$.
As a result, a typical realization for the observed continuum light curve is obtained by
\begin{equation}
\mathbi{\tilde{y}}_{\rm c} = (\mathbi{u_{s}} + \mathbi{\hat s}) + \mathbi{L}(\mathbi{u_q}+\mathbi{\hat q}),
\label{eqn_con}
\end{equation}
where $\mathbi{u_s}$ and $\mathbi{u_q}$ are Gaussian processes with zero mean and covariance matrices $\mathbi{Q}$ and 
$\mathbi{C_q}$, respectively. Note that given ($\sigma_{\rm d}, \tau_{\rm d}, \mathbi{q}$), $\mathbi{\hat s}$
and $\mathbi{\hat q}$ are uniquely determined by Equations~(\ref{eqn_s}) and (\ref{eqn_q}).
In the following analysis, we use $\mathbi{u_s}$ and $\mathbi{u_q}$ as free parameters, which 
are further constrained by additional measured data of broad emission lines.

\section{BLR Modeling}
As mentioned above, \cite{Pancoast2011} developed a Bayesian approach for 
BLR dynamical modeling, which was firstly applied to the RM data of 
Arp 151 (\citealt{Brewer2011b}) and Mrk 50 (\citealt{Pancoast2012}).
\cite{Li2013} carried out an independent implementation of this approach by additionally including 
the non-linear response of the emission lines to the continuum and detrending of light curves. 
\cite{Pancoast2014a} further improved their BLR model by including 
more complicated treatments on anisotropy of line emissions and on cloud kinematics so as to
generate highly asymmetric line profiles. This improved approach was
subsequently applied to several RM objects by \cite{Pancoast2014b}, \cite{Grier2017},
\cite{Pancoast2018}, and {\cb\cite{Williams2018}}.
The BLR models used in this paper are based on \cite{Pancoast2014a} and \cite{Li2013}, but with
several new modifications. Here, for the sake of completeness, we list all the essential details.

The basic scenario of BLR dynamical modeling is that BLRs are composed of a large 
number of discrete, point-like clouds, which orbit around 
the central BH (e.g., \citealt{Netzer1990}). These clouds are exposed to 
the central ionizing source and instantaneously re-radiate emission lines by absorbing 
the ionizing continuum. Due to lacking UV/X-ray monitoring data, we use 
5100~{\AA} fluxes as a surrogate for the ionizing continuum. This may lead to the
non-linear response of BLR cloud emission (\citealt{Gaskell1986, Goad2014}).
{\cb To avoid confusion, hereafter we use the term ``particles''
to represent units of emissions from these BLR clouds.  The effect of 
inverse square decline of the incident continuum flux density is assumed to be implicitly included in the 
radial distribution of BLR particles.}
Below we construct three BLR models $M1$, $M2$, and $M3$. 
The first model $M1$ is the same as \cite{Pancoast2014a}'s model.
$M2$ differs from $M1$ at the prescription of the radial distributions of BLR particles. $M3$ is a two-zone model motivated by 
the theoretical model of \cite{Wang2014b}.
Throughout the calculations, a spherical coordinate frame ($r, \theta, \varphi$) is used
and the BH is placed in the origin.

\subsection{Geometry}
The distribution of BLR particles is assumed to be axisymmetric and follows a flexible  
disk-like geometry, which can yield a variety of shapes with suitable parameters, 
including shells, spheres, and rings. 
The BLR has an inclination angle $\theta_{\rm inc}$ to the observer, which 
is defined by the angle between the line of sight and the symmetric axis of the BLR.
The BLR particles subtend an opening angle $\theta_{\rm opn}$, which is defined by
$\theta_{\rm opn}=\pi/2$ for a spherical BLR and $\theta_{\rm opn}=0$
for a infinitely thin disk-like BLR 
(see the schematic Figure 1 of \citealt{Li2013}). Within the opening angle,
particles are distributed uniformly over $\varphi$-direction.
In $\theta$-direction, particles are distributed with a prescription (\citealt{Pancoast2014a})
\begin{equation}
\theta = \cos^{-1}\left[\cos\theta_{\rm opn} + (1-\cos\theta_{\rm opn})\times U^{\gamma}\right],
\end{equation}
where $U$ is a random number from a uniform distribution between 0 and 1 and 
$\gamma$ is a free parameter that controls the extent to which particles are clustered along the outer face 
of the BLR disk.

We use two types of radial distribution for BLR particles as follows. 
\begin{itemize}
 \item $M1$: The radial distribution is parameterized by a 
Gamma distribution, same as in \citealt{Pancoast2014a}.
Specifically, the radial location of a particle is assigned by
\begin{equation}
 r=F\mu + (1-F)\mathscr{R},
\label{eqn_radial}
\end{equation}
where $\mathscr{R}$ is a random number drawn from the Gamma distribution with a mean $\mu$ 
and a standard deviation $\beta\mu$\footnote{If the Gamma
function is parameterized with a shape parameter ($a$) and a scale parameter ($s$),
they obey the relations $a = 1/\beta^2$ and $s=\beta^2\mu$.}, and 
$F$ is a fraction to account for the possibility that within an inner edge ($F\mu$), 
clouds are completely ionized so that they do not reverberate to the continuum. 

\item $M2$: The radial distribution is parameterized by a double power law (\citealt{Stern2015}) as
\begin{equation}
f(r) \propto \left\{\begin{array}{ll}
        r^{\alpha}, & {\rm for}~F_{\rm in}\leqslant r/R_0 \leqslant 1,\\
        r^{-\alpha},& {\rm for}~1\leqslant r/R_0 \leqslant F_{\rm out},
       \end{array}\right.
\end{equation}
where $\alpha$ is the slope of the power law, $R_0$ is the characteristic radius, and $F_{\rm in}$ and $F_{\rm out}$
are fractions to describe the inner and outer radius.
\end{itemize}
Hereafter, we also denote the BLR model with the Gamma distribution as $M1$ and with the power-law distribution 
as $M2$.

\subsection{Emissivity}
We assume that the ionizing continuum is isotropic, and is proportional to 
the optical 5100~{\AA} continuum. Self-shadowing among clouds is not 
considered for the present simple modeling. 
As in \cite{Li2013}, we relax the usual assumption of 
linear responses of emission lines to the continuum
and adopt a power-law index $\delta$ to describe the non-linearity
as
\begin{equation}
\epsilon(t)\propto f_{\rm c}^{1+\delta}(t-\tau), 
\end{equation}
where $\epsilon$ is the emissivity of the particle at time $t$ irradiated by 
the ionizing continuum with a flux of $f_{\rm c}$ at time $t-\tau$.

To account for the possibility that BLR clouds are optically thick so that 
their emission is anisotropic, we use a simple parameterization 
by assigning a weight to each particle as (\citealt{Blandford1982}) 
\begin{equation}
w = \frac{1}{2} + \kappa \cos\phi,
\label{eqn_wei}
\end{equation}
where $\kappa$ is a free factor in the range of [-1/2, 1/2] and $\phi$ is
the angle between the observer's and particle's line of sight to the central ionizing 
source. 

It is possible that the particles below the equatorial plane are partially obscured 
by some material in the equatorial plane. As in \cite{Pancoast2014a}, we use a 
parameter $\xi$ to describe the transparency of this equatorial material.
For $\xi\rightarrow0$, the entire half of the BLR below the equatorial plane is
obscured; whereas for $\xi\rightarrow1$, the half becomes transparent.

\subsection{Dynamics}
The motion of particles is assumed to be fully dominated by the gravity of the central BH. 
Following \cite{Pancoast2014a}, three kinematic components are considered: bound elliptical orbits, and bound and unbound inflow or outflow. 
The fraction of bound elliptical orbits is described a by parameter $f_{\rm ellip}$ and the remaining fraction $1-f_{\rm ellip}$ 
of BLR particles is thus either inflowing or outflowing. A parameter $f_{\rm flow}$ is used to determine 
whether BLR particles are inflowing ($0<f_{\rm flow}<0.5$) or outflowing ($0.5<f_{\rm flow}<1$). 

Velocities of particles are firstly assigned in the particles' orbital planes and then converted into real three-dimension velocities
through coordinate rotations. For bound elliptical orbits, radial and tangential velocities are drawn from Gaussian distributions
centered around the point $(v_r, v_\phi)=(0, v_{\rm circ})$ of an ellipse in the $v_r-v_\phi$ plane (see Figure 2 in 
\citealt{Pancoast2014a}), where $v_{\rm circ}=\sqrt{GM_\bullet/r}$. The ellipse has a semiminor axis $v_{\rm circ}$ in the $v_\phi$-direction
and a semimajor axis $\sqrt{2}v_{\rm circ}$ in the $v_r$-direction. The widths of Gaussian distributions for radial and tangential velocities
are controlled by parameters $\sigma_{\rho, \rm circ}$ and $\sigma_{\Theta, \rm circ}$, respectively, where $\rho$ and $\Theta$ are 
the radial and angular coordinates in the $v_r-v_\phi$ plane.

For inflowing or outflowing particles, velocities are assigned same as for elliptical orbits, except that the Gaussian 
distributions are centered around points $(v_r, v_\phi)=(\pm\sqrt{2}v_{\rm circ}, 0)$ in the $v_r-v_\phi$ plane,
where ``$+$'' corresponds to outflow and ``$-$'' corresponds to inflow.
In addition, the Gaussian distributions are allowed to rotate along the ellipse by an angle $\theta_{\rm e}$ considering that
real clouds may have a combination of Keplerian and inflow/outflow motion.
When $\theta_e=0$, inflowing or outflowing velocities are centered around the escape velocity $v_r=\pm\sqrt{2}v_{\rm circ}$.
As $\theta_e\rightarrow90^\circ$, inflowing or outflowing particles approach the same motion as the elliptical orbits.
 
Macroturbulence is included by adding a random velocity to the light-of-sight velocity of particles as (\citealt{Pancoast2014a})
\begin{equation}
v_{\rm turb}=\mathcal{N}(0, \sigma_{\rm turb})v_{\rm circ},
\end{equation}
where $\mathcal{N}(0, \sigma_{\rm turb})$ is random number drawn from a Gaussian distribution with a zero mean and standard deviation $\sigma_{\rm turb}$.

\subsection{A Two-zone BLR Model $M3$}
At high accretion rates, accretion disks are usually categorized into slim disk regime (\citealt{Abramowicz1988}), which forms
a geometrically thick funnel in the inner disks due to the strong radiation pressure.
\cite{Wang2014b} showed that the self-shadowing effect of such a funnel feature produces anisotropy 
of the ionizing radiation field that leads to two distinct BLR regions. 
Mrk 142 was identified to be an SEAMBH with a dimensionless accretion rate of
$\mathscr{\dot M}=45$ {\cb using the BH mass derived from cross-correlation function (CCF) analysis and 
an assumed inclination angle of $\cos\theta_{\rm inc}=0.75$}
(\citealt{Du2016a}), where $\mathscr{\dot M}=\dot M c^2/L_{\rm Edd}$, $\dot M$ is mass accretion rate,
and $L_{\rm Edd}$ is the Eddington luminosity. Motivated by the above scenario,
we construct a two-zone BLR model. Figure~\ref{fig_sch} shows a schematic of the two-zone BLR geometry for Mrk 142.
Zone I is ionized by radiation emitted within the funnel whereas zone II is ionized by radiation emitted outside
the funnel. For simplicity, we assume that the ionizing emissions received by zone I and zone II
are correlated and we thereby apply the observed 5100~{\AA} continuum light curve for the both regions.
In addition, we neglect the obscuration of zone I to zone II.

The configuration of this two-zone model is set as follows.
The structure and dynamics of zone I are described with the same parameters as in model $M1$ in Section 4.1. 
For zone~II, the radial distribution of BLR particles follows a Gamma distribution but with distinct parameters.
Meanwhile, zone~I also has new dynamical parameters $f_{\rm ellip, I}$ and $f_{\rm flow, I}$. 
Hereafter, we denote this two-zone model as $M3$. In a nutshell, compared to model $M1$, $M3$ has new additional 
parameters ($\mu_{\rm I}$, $\beta_{\rm I}$, $F_{\rm I}$, $\theta_{\rm opn, I}$, $\rho_{\rm I}$, $f_{\rm ellip, I}$, and $f_{\rm flow, I}$).
The meanings of these parameters are also explained in Table~\ref{tab_par}.

\subsection{Different Long-term Trends of Continuum and Emission Line}
\label{sec_trend}
There is incident detection in previous RM observations that the variations of continuum and emission 
line undergo different long-term (compared with RM timescales) secular trends (e.g., \citealt{Denney2010, Li2013, Peterson2014b}), 
which are irrelevant to RM analysis and therefore should be appropriately accounted. A low-order polynomial 
was usually used to detrend the light curves of continuum and emission line (\citealt{Welsh1999}), which generally 
leads to improvements in the RM analysis.

In the present framework, we use a linear polynomial to model the difference in the long-term trends of continuum
and emission line. We add this linear trend to the reconstructed continuum light curve so that the new  
light curve has the same secular trend as that of emission line. To keep the mean flux of the continuum 
light curve unchanged, only a free parameter is needed to delineate the slope the linear polynomial.
There are no apparently different trends by visual inspection in the light curves of Mrk 142, therefore we
do not include this procedure in our calculations. However, there are indeed a few objects in our monitored sample 
showing different long-term trends. We describe the procedure for including different long-term trends
here for the sake of completeness. We stress that here the linear trend 
serves the purpose of accounting for the different long-term trends in continuum and emission line, distinguished from 
the trend defined in Section 3.1, which only refers to the continuum itself. 

\section{Bayesian Framework}
\subsection{Formulations}
It is now trivial to calculate the intrinsic emission line profile at time~$t$ and 
velocity $v$ by summing up the emissions from BLR particles with a line-of-sight velocity $v$
\begin{eqnarray}
f_{l,\rm int}(v, t) & = &\sum_i\epsilon_i(v,t) 
    =  A\sum_i \delta(v-u_i)w_i f^{1+\gamma}_c(t-\tau_i),
\label{eqn_rev}
\end{eqnarray}
where $\delta(x)$ is the Dirac function, 
$A$ is the response coefficient, and $w_i$, $u_i$, $r_i$, and $\tau_i$ 
are, respectively, the weight of emissivity
(given by Equation (\ref{eqn_wei})), the line-of-sight velocity, the distance to the central source, and 
the time-lag of re-radiation from the $i$-th particle. The continuum flux $f_c$ includes the different 
long-term trends in continuum and emission, as described in Section~\ref{sec_trend}. 
Here the subscript ``int'' refers to intrinsic line profile, to distinguish from the observed 
line profile, which suffers additional broadening due to the seeing and instrument effects.
The so-called transfer function reads
\begin{equation}
\Psi_{\rm int}(v, \tau) = A\sum_i w_i\delta(v-u_i)\delta(\tau-\tau_i).
\end{equation}
This simplifies Equation (\ref{eqn_rev}) into a generalized integral form for RM
with a non-linear response,
\begin{equation}
f_{l, \rm int}(v, t) =  \int \Psi_{\rm int}(v,\tau)f_c^{1+\gamma}(t-\tau)d\tau.
\label{eqn_rm}
\end{equation}
For {\it long} time series, manipulating time averaging upon both sides of the above equation 
yields the delay integral of $\Psi(v, t)$
\begin{equation}
\Psi_{\rm int}(v) = \int \Psi_{\rm int}(v, \tau)d\tau =
\frac{\langle f_{l,\rm int}(v, t)\rangle}{\langle f_c(t)\rangle},
\label{eqn_psi_int}
\end{equation}
where the angle brackets denote time average. The velocity integral of $\Psi(v,\tau)$,
\begin{equation}
\Psi(\tau) = \Psi_{\rm int}(\tau) = \int \Psi_{\rm int}(v, \tau) dv,
\label{eqn_psi_tau}
\end{equation}
yields the usual velocity-unresolved delay map.

To mock real observations, we need to take into account line broadening caused by the seeing and instruments.
The observed line profile can be deemed to be a convolution between the predicted intrinsic 
line profile and line-broadening function.
Equation~(\ref{eqn_rm}) is recasted into
\begin{eqnarray}
f_l(v, t) & = &f_{l, \rm int}(v, t)\otimes \xi(v, t)\nonumber\\
          & = &\int \Psi_{\rm int}(v,\tau)\otimes \xi(v, t) f_c^{1+\gamma}(t-\tau)d\tau,
\label{eqn_line_obs}
\end{eqnarray}
where ``$\otimes$'' denotes a convolution operation over velocity axis and $\xi(v, t)$ is the 
line-broadening function, which generally depends on the instrument and seeing conditions. 
We again manipulate time 
averaging upon both sides of the above equation and note that $\xi(v, t)$ and $f_c(t)$ are 
usually uncorrelated. For {\it long} time series, we have
\begin{equation}
\langle f_l(v, t) \rangle = \langle f_c(t) \rangle \int \Psi_{\rm int}(v, \tau)\otimes 
\langle \xi(v, t) \rangle d\tau.
\end{equation}
If we denote
\begin{equation}
\Psi(v) = \int \Psi_{\rm int}(v, \tau)\otimes 
\langle \xi(v, t) \rangle d\tau,
\end{equation}
we obtain exactly the same form for $\Psi(v)$ as Equation~(\ref{eqn_psi_int}),
\begin{equation}
\Psi(v) = \frac{\langle f_l(v, t)\rangle}{\langle f_c(t) \rangle} \propto \langle f_l(v, t)\rangle.
\label{eqn_psi}
\end{equation}
This implies that the delay integral of (broadened) transfer function has the same shape 
as the mean observed profile of the emission line (\citealt{Blandford1982, Perry1994}).
{\cb Unless stated otherwise, transfer functions shown in figures throughout the paper
by default include the broadening effect, namely, convolution with the line-broadening function.}

A set of measurements for an emission line in real observations is a sum of the predicted line 
profiles and measurement noises. Written in a concise form of tensors
\begin{equation}
\mathbi{y}_l = \mathbi{f}_l + \mathbi{n}_l = \mathbi{f}_{l,\rm int}\otimes\boldsymbol{\xi} + \mathbi{n}_l,
\end{equation}
where $\mathbi{n}_l$ are measurement noises. For simplicity,
we parameterize the line-broadening function $\boldsymbol{\xi}$ by a Gaussian and adopt the 
dispersion from the mean value derived by \citetalias{Du2016b}. The value of the dispersion is fixed 
throughout the period of the RM data (see Table~\ref{tab_sample}).
Again, we assume that the measurement noises $\mathbi{n}_l$ are Gaussian and 
uncorrelated along both wavelength and time axes.
This results in a Gaussian likelihood probability for $\mathbi{y}_l$ as
\begin{equation}
P_l(\mathbi{y}_l|\boldsymbol{\Theta})=\prod_{ij}\frac{1}{\sqrt{2\pi}\sigma_{ij}}
\exp\left\lgroup-\frac{(y_{l, ij}-f_{l, ij})^2}{2\sigma_{ij}^2}\right\rgroup,
\end{equation}
where $\boldsymbol{\Theta}$ denotes the whole set of involved model parameters listed in Table~\ref{tab_par}, 
$i$ and $j$ represent the epoch and wavelength bin, and $\sigma_{ij}$ is the measurement noise.

\begin{figure*}[ht!]
\centering
\includegraphics[width=0.85\textwidth]{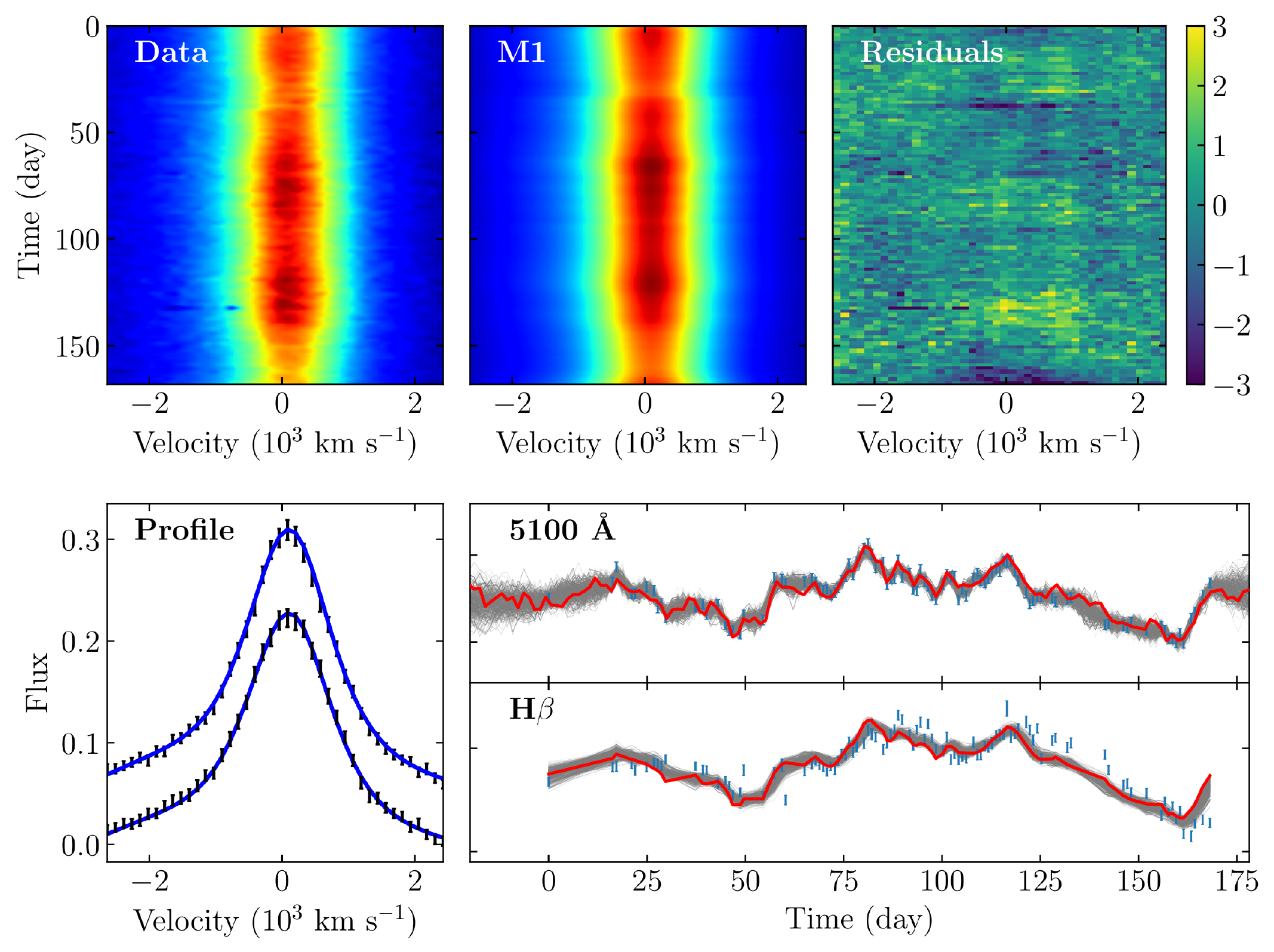}
\caption{Fits to the RM data of Mrk 142 with BLR model $M1$. Top three panels show the observed H$\beta$ spectral 
time series, a model fit, and the residuals between the observed data and model fit, which are normalized 
by the square of the measurement errors. Bottom left panel
shows H$\beta$ profiles at two selected epochs, superposed upon the model fits with blue solid lines.
Bottom right panels show the time series of the 5100~{\AA} continuum and H$\beta$ fluxes. Red line shows 
the best recovered time series and light grey lines represent random reconstructions. The observations are 
set to start at day zero.}
\label{fig_m1}
\end{figure*}

\begin{figure*}[ht!]
\centering
\includegraphics[width=0.85\textwidth]{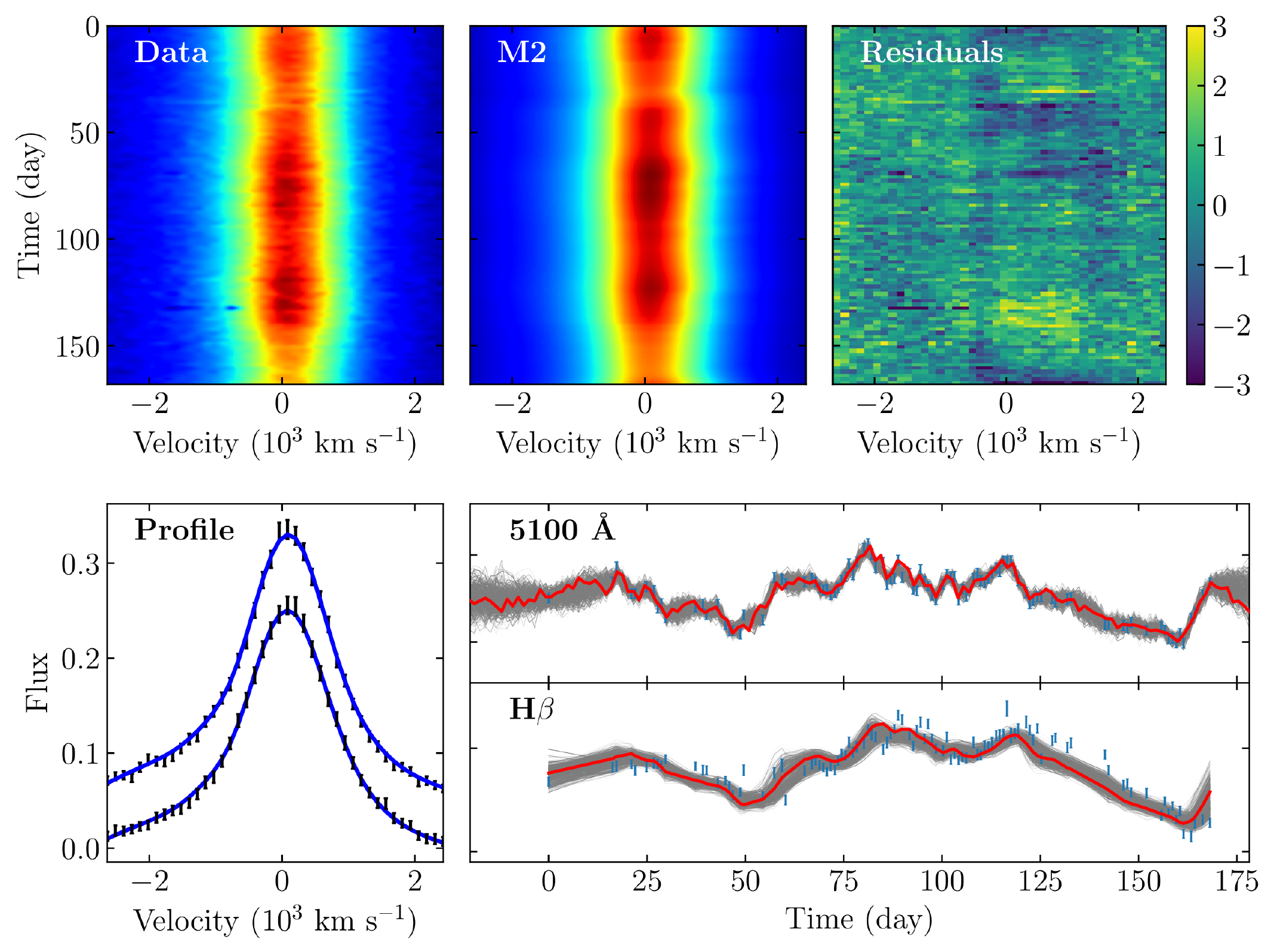}
\caption{Same as Figure \ref{fig_m1} but for BLR model $M2$.}
\label{fig_m2}
\end{figure*}

\begin{figure*}[ht!]
\centering
\includegraphics[width=0.85\textwidth]{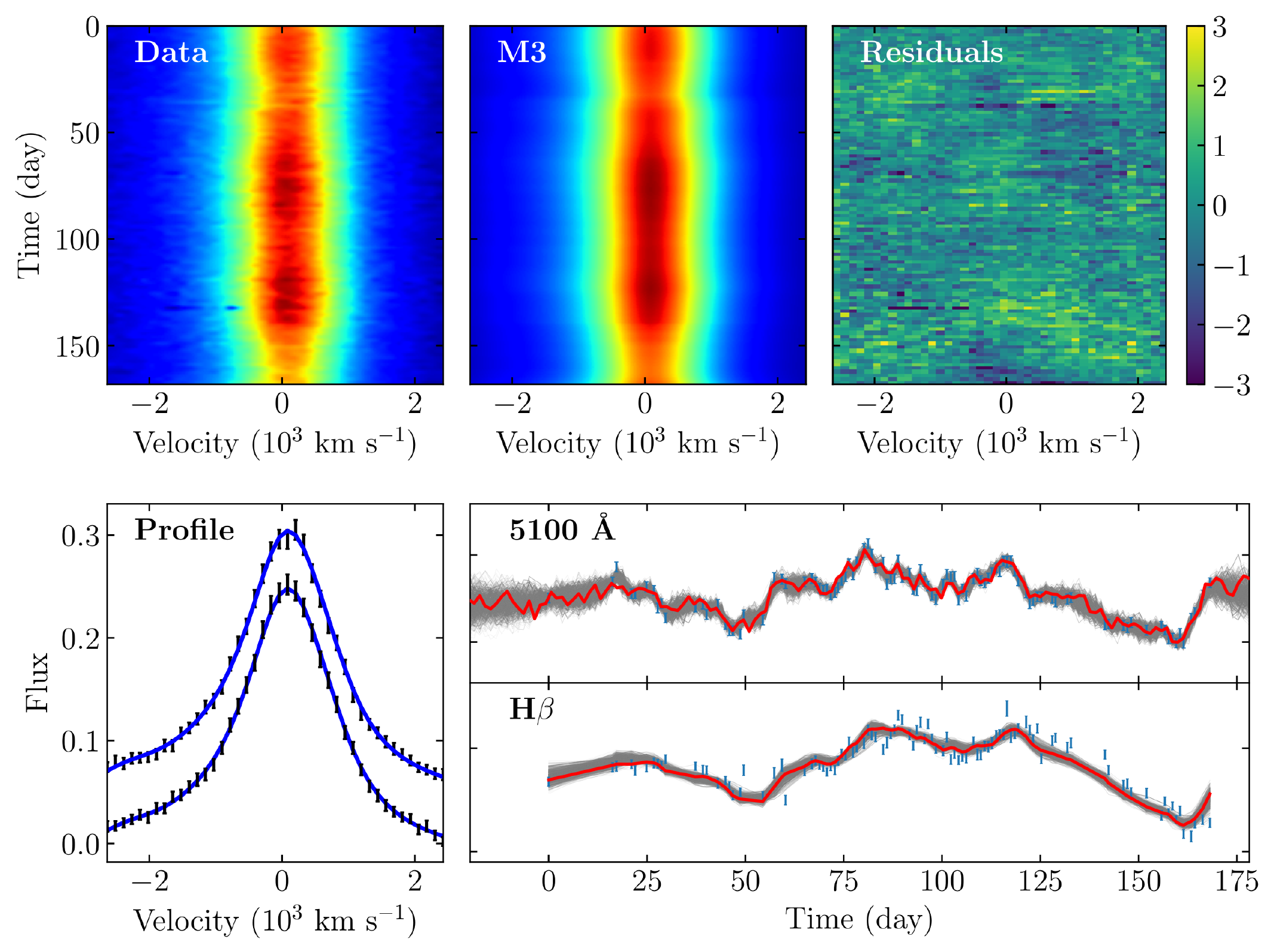}
\caption{Same as Figure \ref{fig_m1} but for BLR model $M3$.}
\label{fig_m3}
\end{figure*}

\subsection{Bayesian Inference}

The RM data ($\mathbi{D}$) at hand are the time series of continuum $\mathbi{y}_{\rm c}$ and emission line $\mathbi{y}_l$ 
and their respective associated measurement errors. We first use Equation~(\ref{eqn_con}) to generate realizations for 
the continuum time series, which are then used as input for deriving time series of emission line with
Equations~(\ref{eqn_rev}) and (\ref{eqn_line_obs}). In this regard, the continuum data are treated as a prior for BLR modeling
(A. Pancoast, private communications; \citealt{Pancoast2011, Pancoast2014a}).
From this paradigm, the likelihood probability for $\mathbi{D}$ is thereby 
\begin{equation}
P(\mathbi{D}|\boldsymbol{\Theta}) = P(\mathbi{y}_l|\boldsymbol{\Theta}).
\label{eqn_like}
\end{equation}
According to the Bayes' theorem, the posterior probability distribution for 
the parameter set $\boldsymbol{\Theta}$ is 
\begin{equation}
 P(\boldsymbol{\Theta}|\mathbi{D}) = \frac{P(\boldsymbol{\Theta})P(\mathbi{D}|\boldsymbol{\Theta})}{P(\mathbi{D})},
 \label{eqn_post}
\end{equation}
where $P(\boldsymbol{\Theta})$ is the prior for the parameter set and $P(\mathbi{D})$ is 
the Bayesian evidence that just plays the role of normalization factor and important for model selection. 
The priors for $\mathbi{u_s}$ and $\mathbi{u_q}$ are Gaussian, as described in 
Section~\ref{sec_con}. The priors for the other parameters are listed in Table~\ref{tab_par}, which are assigned 
following the convention that for parameters whose typical value ranges are known, a uniform
prior is assigned; otherwise, if the parameter information is
completely unknown, a logarithmic prior is assigned (\citealt{Sivia2006}, Chapter~5).
For all the priors, we set a reasonably broad but still finite
range to avoid the posterior impropriety.

We include an extra noise parameter $\sigma_{{\rm s}, l}$, added in square to the measurement noises of emission line data. 
This is based on two considerations: the present model for BLRs is simple so that it unlikely fits all the features of data; 
on the other hand, there are probably additional noises beyond the known measurement uncertainties.
Another additional advantage of using an extra noise parameter is that this provides a very useful
annealing operation that benefits the convergences of Markov chains (see below) when the initial 
parameter values are far from the best-fitting values.

In Table \ref{tab_par}, we list the overall free parameters for continuum and BLR modeling. 
We use 200 points to describe continuum light curve, leading to a total of
hundreds of free parameters. Meanwhile, there are strong correlations among 
BLR parameters, such as inclination and BH mass. This requires sophisticated algorithms 
that can handle massive and highly correlated parameters.
We use the Markov Chain Monte Carlo (MCMC) method to construct samples from the posterior 
distribution and determine the best-fitting estimate for the parameters. We employ the diffusive 
nested sampling (DNS) algorithm proposed by \cite{Brewer2011a} to generate the Markov chains. 
The DNS algorithm is effective
at exploring multimodal distributions and strong correlations between parameters.
It also allows us to calculate the Bayesian evidence, which can be used for subsequent 
model selection. Moreover, the DNS algorithm is inherently parallel and is easy to implement
on parallel computing interfaces. We write our own DNS code in C language using the
standardized Message Passing Interface (MPI) so that the code is portable to 
a wide range of supercomputer clusters without any reliance on special features of proprietary compilers.
We develop a code named \texttt{BRAINS} to implement the above BLR dynamical modeling and 
Bayesian inference, which is publicly available at \texttt{https://github.com/LiyrAstroph/BRAINS}.
Unless stated otherwise, throughout the calculations, {\cb the best estimates for the parameters are
taken to be the median values of their posterior distributions and the
uncertainties are determined from the 68.3\% confidence intervals.}

\begin{figure}
\centering 
\includegraphics[width=0.48\textwidth]{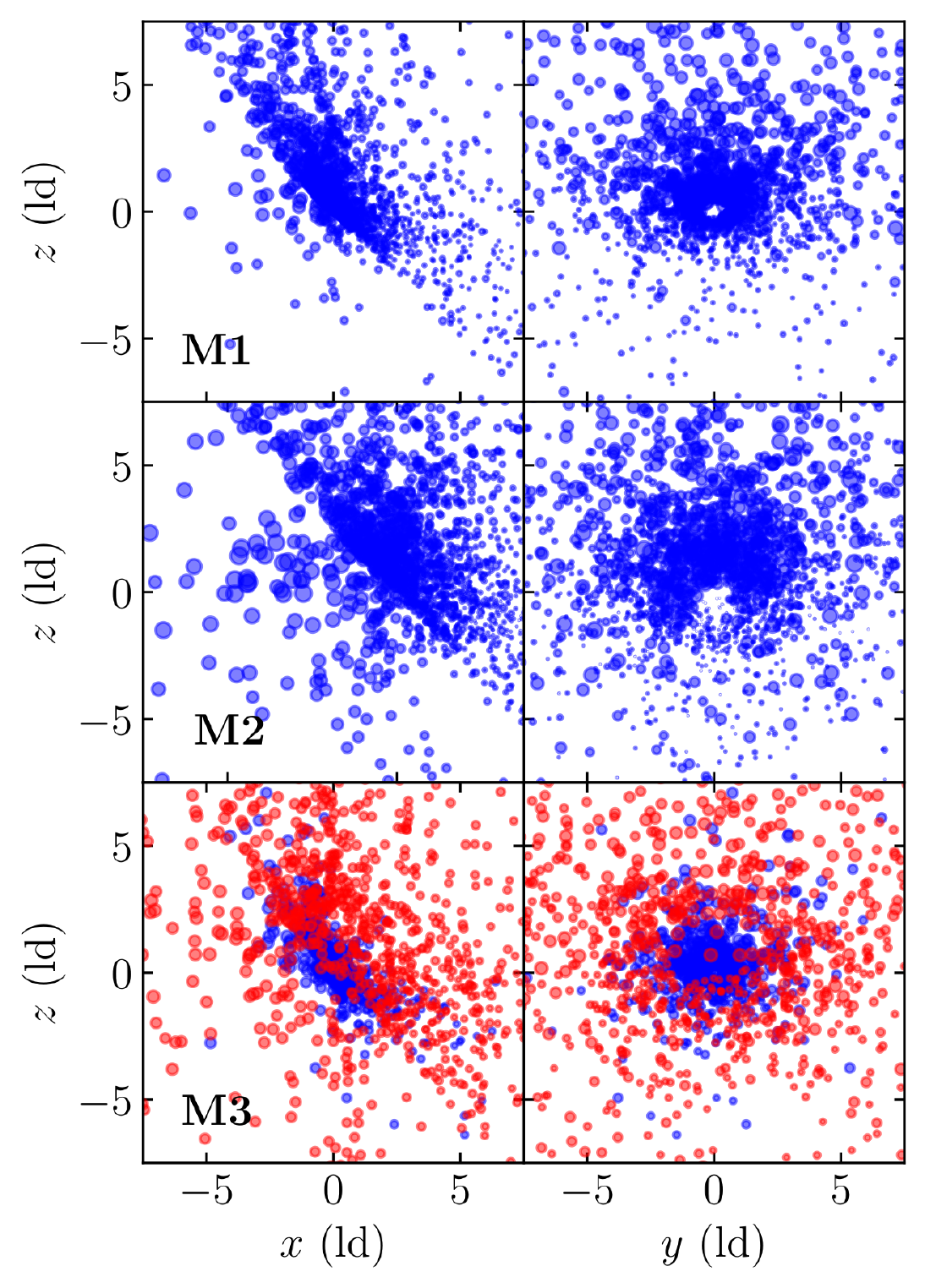}
\caption{\cb Examples of inferred geometry of the BLR for the three models from top to bottom. Each point represents 
a BLR particle and its size is in proportional to the weight of emissivity in Equation (15). For model $M3$, 
red and blue points correspond to BLR particles from zone I and zone II, respectively.}
\label{fig_cloud}
\end{figure}

\begin{deluxetable}{cccc}
\tabletypesize{\footnotesize}
\tablecolumns{4}
\tablewidth{0.45\textwidth}
\tablecaption{Comparison of the three BLR models $M1$, $M2$, and $M3$.\label{tab_comp}}
\tablehead{
  \colhead{~~~~~~~~~~~~~~~~~}
& \colhead{~~~~~~~~~~~$M1$~~~~~~~~~~~}
& \colhead{~~~~~~~~~~~$M2$~~~~~~~~~~~}
& \colhead{~~~~~~~~~~~$M3$~~~~~~~~~~~}
}
\startdata
$\ln\mathcal{L}_{\rm max}$ & 0   & -280   &  164\\
BIC  &  0    & 288  & -105\\
AIC &  0   & 562 & -313\\
$\log K$ & 0 & -118 & 68
\enddata
\tablecomments{\cb $K$ is the Bayes factor and all values are given with respective to model $M1$.}
\end{deluxetable}
\begin{figure*}[th!]
\centering
\includegraphics[width=0.8\textwidth]{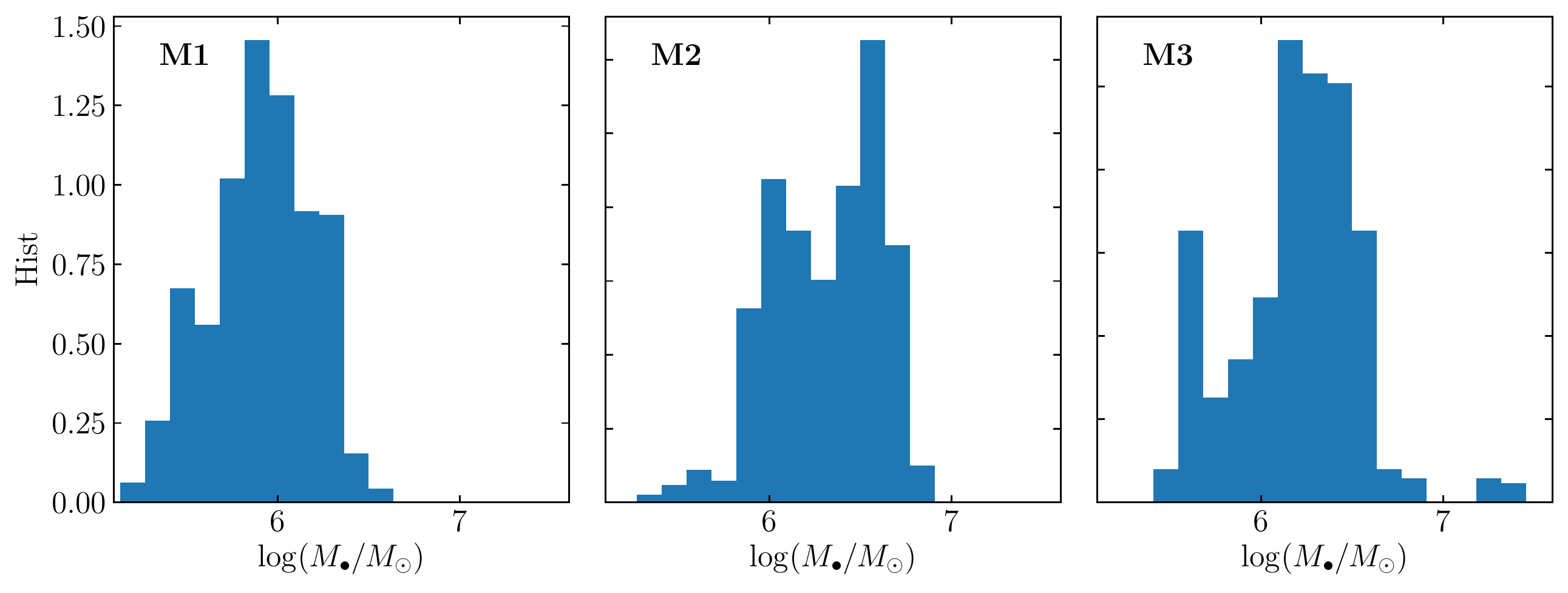}
\caption{Posterior distributions of BH mass obtained with BLR models $M1$, $M2$, and $M3$.}
\label{fig_mass}
\end{figure*}

\section{Results}

\begin{figure*}[th!]
\centering
\includegraphics[width=0.8\textwidth]{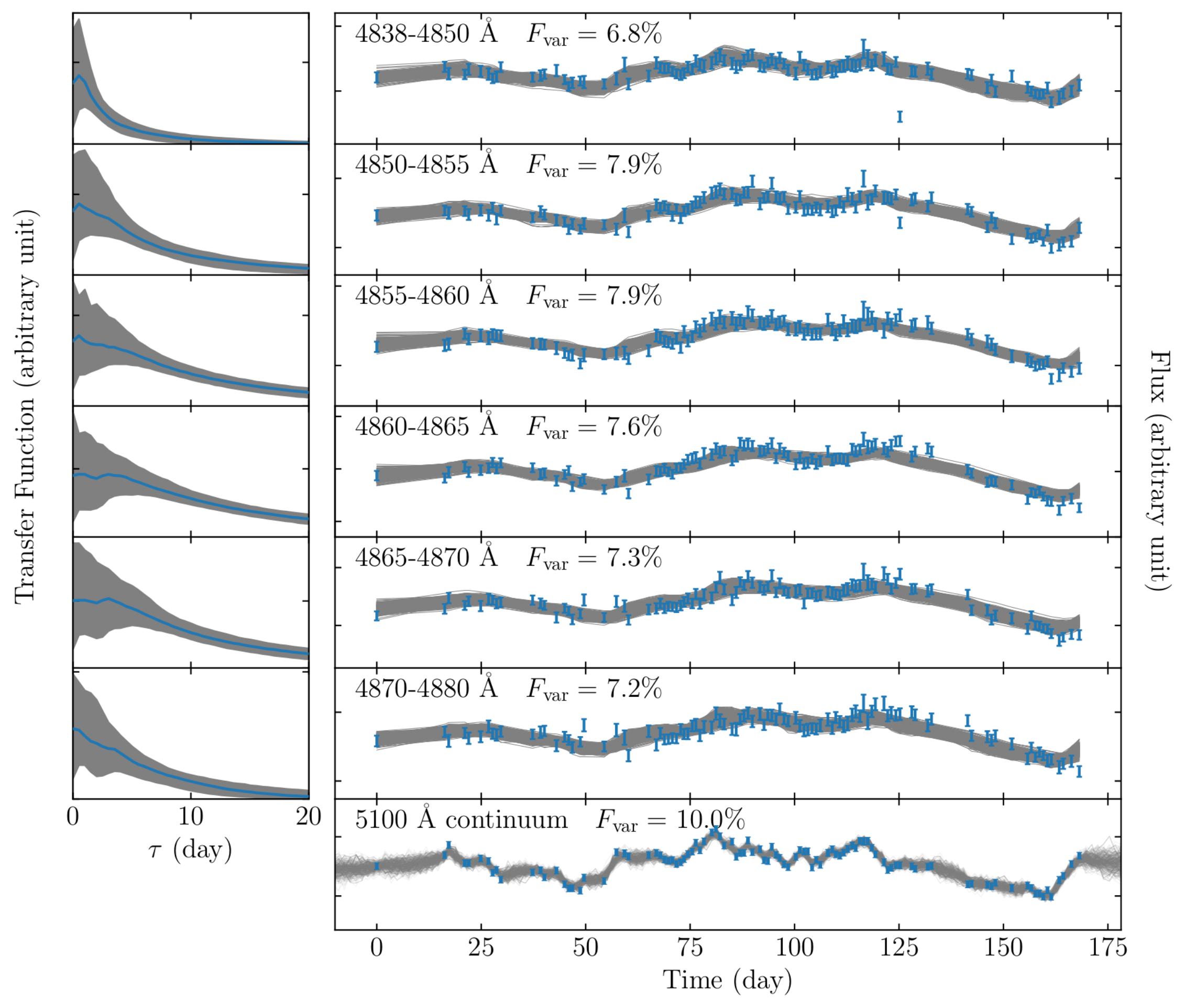}
\caption{(Left panels) The obtained transfer functions based on model $M3$ at selected wavelength bins marked in right panels .
Grey shaded areas represent the 1$\sigma$ error band. (Right panels) The reconstructed light curves. Each thin grey line represents one random 
reconstruction. Points with errorbars are the observed data. The variability characteristic $F_{\rm var}$ of each observed light curve is also
shown.}
\label{fig_tran}
\end{figure*}

\subsection{Overview}
Figures~\ref{fig_m1}-\ref{fig_m3} show the fitting results to the RM data of Mrk~142 with BLR models $M1$, 
$M2$, and $M3$, respectively. In each figure, the top three panels plot the observed H$\beta$ spectral time series, an exemplary
model fit, and the residuals between the data and the fit, respectively. The bottom panels plot 
the recovered H$\beta$ profiles at two selected epochs and the reconstructed light curves of the continuum and H$\beta$ fluxes. 
The three models can generally well reproduce the continuum and H$\beta$ flux light curves, but differ 
with varying degrees of success in reproducing the detailed H$\beta$ spectral time series.
The obtained residuals between the data and model fit illustrate that the fits of $M2$ model
have systematic deviations around $\sim-500~{\rm km~s^{-1}}$ and $\pm1500~{\rm km~s^{-1}}$ of H$\beta$ profiles.
Models $M1$ and $M3$ give similar fitting to the data, with $M3$ slightly better by visual inspection.
Figure~\ref{fig_cloud} shows examples of inferred geometry of the BLR for the three models.
In Figure~\ref{fig_mass}, we plot the posterior distributions of BH mass obtained by the three models.
The best inferred BH mass is {\cb $\log(M_\bullet/M_\odot)=5.90_{-0.31}^{+0.31}$, $6.34_{-0.25}^{+0.37}$, and $6.23_{-0.45}^{+0.23}$} for 
$M1$, $M2$, and $M3$, respectively. {\cb These values are consistent with each other to within uncertainties.
The inferred values for the major parameters of the three models are tabulated in Appendix A.}

\begin{figure*}[th!]
\centering
\includegraphics[width=0.8\textwidth]{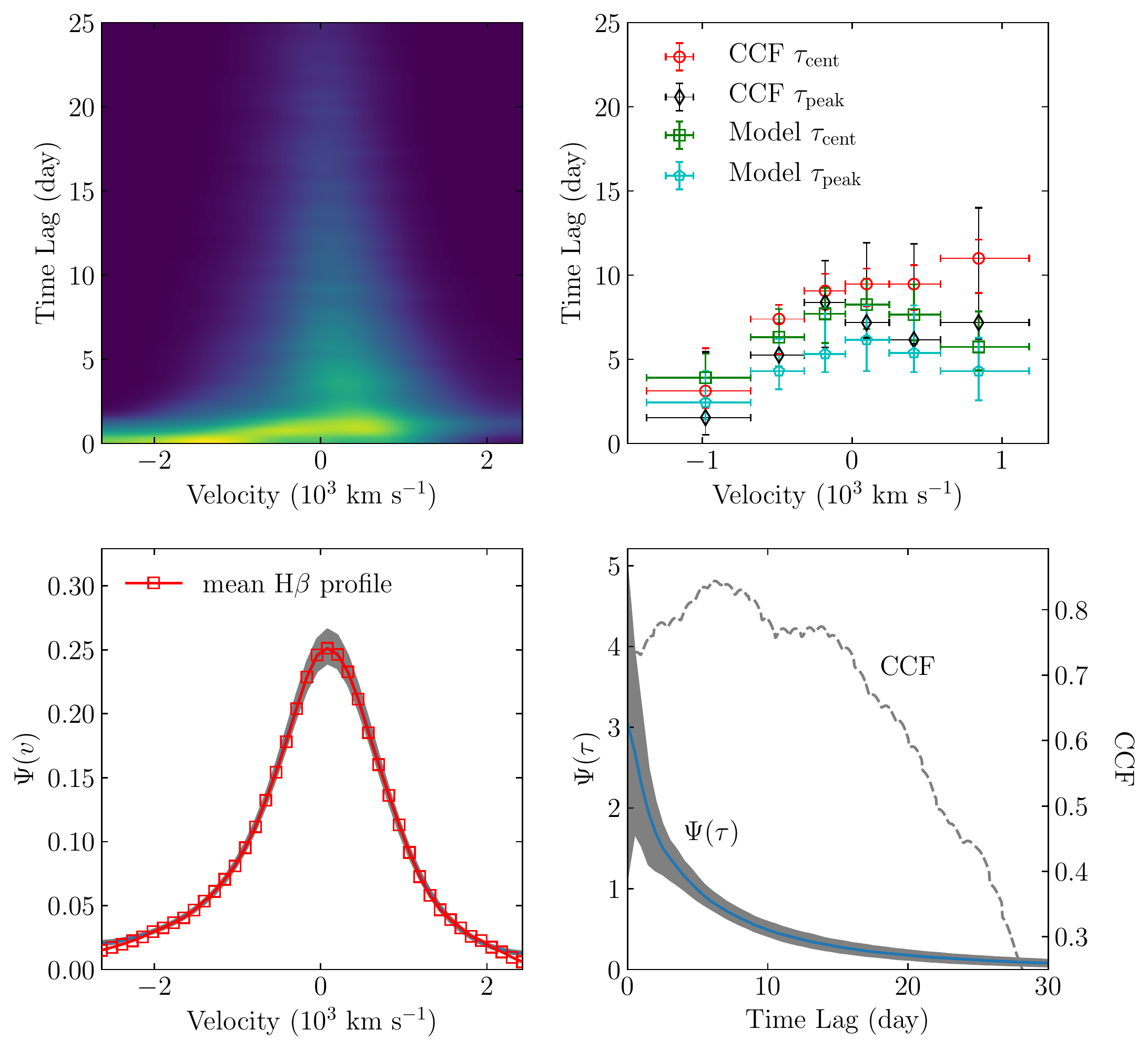}
\caption{(Top left) An example of transfer function obtained using model $M3$. (Top right) Comparison of velocity-binned
time lags obtained from model $M3$ and CCF analysis. 
(Bottom left) Delay integral of transfer function $\Psi(v)$, 
superposed on the observed, scaled mean H$\beta$ profile. Shaded areas represent the 1$\sigma$ error band. 
(Bottom right) Velocity integral of transfer function $\Psi(\tau)$. Shaded areas represent the 1$\sigma$ error band. 
Dashed line represents the CCF between the observed light curves of continuum and H$\beta$ fluxes.
}
\label{fig_ccf}
\end{figure*}

We use the standard approaches of model comparison to determine the best model.
In Table \ref{tab_comp}, we calculate the maximum likelihood ($\ln\mathcal{L}_{\rm max}$), 
the Bayesian information criterion (BIC%
\footnote{The BIC is defined by (\citealt{Schwarz1978})
\begin{equation}
{\rm BIC} = k\ln n - 2\ln \mathcal L_{\rm max},
\end{equation}
where $k$ is the number of model parameters, $n$ is the sample size, and $\mathcal{L}_{\rm max}$ is the maximum value of 
the likelihood function.
}), 
the Akaike information criterion (AIC%
\footnote{The AIC is defined by (\citealt{Akaike1973})
\begin{equation}
{\rm AIC} = 2k - 2\ln \mathcal L_{\rm max},
\end{equation}
where $k$ is the number of model parameters. The original form of AIC is only strictly valid asymptotically. \cite{Hurvick1989}
proposed a correction to AIC for finite sample size, defined as 
\begin{equation}
{\rm AIC} = 2k - 2\ln \mathcal L_{\rm max} + \frac{2k(k+1)}{n-k-1},
\end{equation}
where $n$ is the sample size. We use this corrected AIC in our calculations.
}),
and the Bayes factor %
\footnote{Bayes factor is defined by the ratio of the posterior probabilities (\citealt{Sivia2006}). For two models, say $M1$ and $M2$
with equal priors, the Bayes factor is equal to the ratio of the corresponding Bayesian evidence 
\begin{equation}
K=\frac{P(M2|\mathbi{D})}{P(M1|\mathbi{D})} = \frac{P(\mathbi{D}|M2)}{P(\mathbi{D}|M1)}.
\end{equation}
}
for the three models. The best model is chosen to be the one that maximizes the $\mathcal{L}_{\rm max}$ and Bayes factor and 
minimizes the AIC and BIC.  
All the approaches rank model $M3$ as the best model for fitting the RM data of Mrk 142.
In the following, we study the BLR structure in Mrk 142 based only on the results from model M3.

Figure~\ref{fig_tran} plots the obtained transfer functions and the reconstructed H$\beta$ light curves 
over selected velocity bins. At each velocity bins, the transfer function peaks at zero lag and then gradually 
decreases, typical features seen in transfer functions of inclined disk-like BLRs (e.g., \citealt{Goad1996, Pancoast2014b}). 
We note that the variability characteristic%
\footnote{The variability characteristic of a light curve is defined to be (\citealt{Rodriguez1997})
\begin{equation}
F_{\rm var} = \frac{(\sigma^2-\Delta^2)^{1/2}}{\langle F \rangle},
\end{equation}
where $\langle F \rangle$ is the averaged flux and 
\begin{equation}
 \sigma^2 = \frac{1}{N-1}\sum_{i=1}^{N}(F_i-\langle F \rangle)^2,~~~~\Delta^2 = \frac{1}{N}\sum_{i=1}^{N}\Delta_i^2,
\end{equation}
where $N$ is the number of points and $\Delta_i$ is the uncertainty on the flux $F_i$.
}
at all the velocity bins is generally small ($F_{\rm var}\sim7\%$). Compared with AGNs at sub-Eddington accretion rates,
low variation amplitudes are a major challenge for monitoring SEAMBH objects (\citealt{Rakshit2017}).

\begin{figure*}[th!]
\centering
\includegraphics[width=0.85\textwidth]{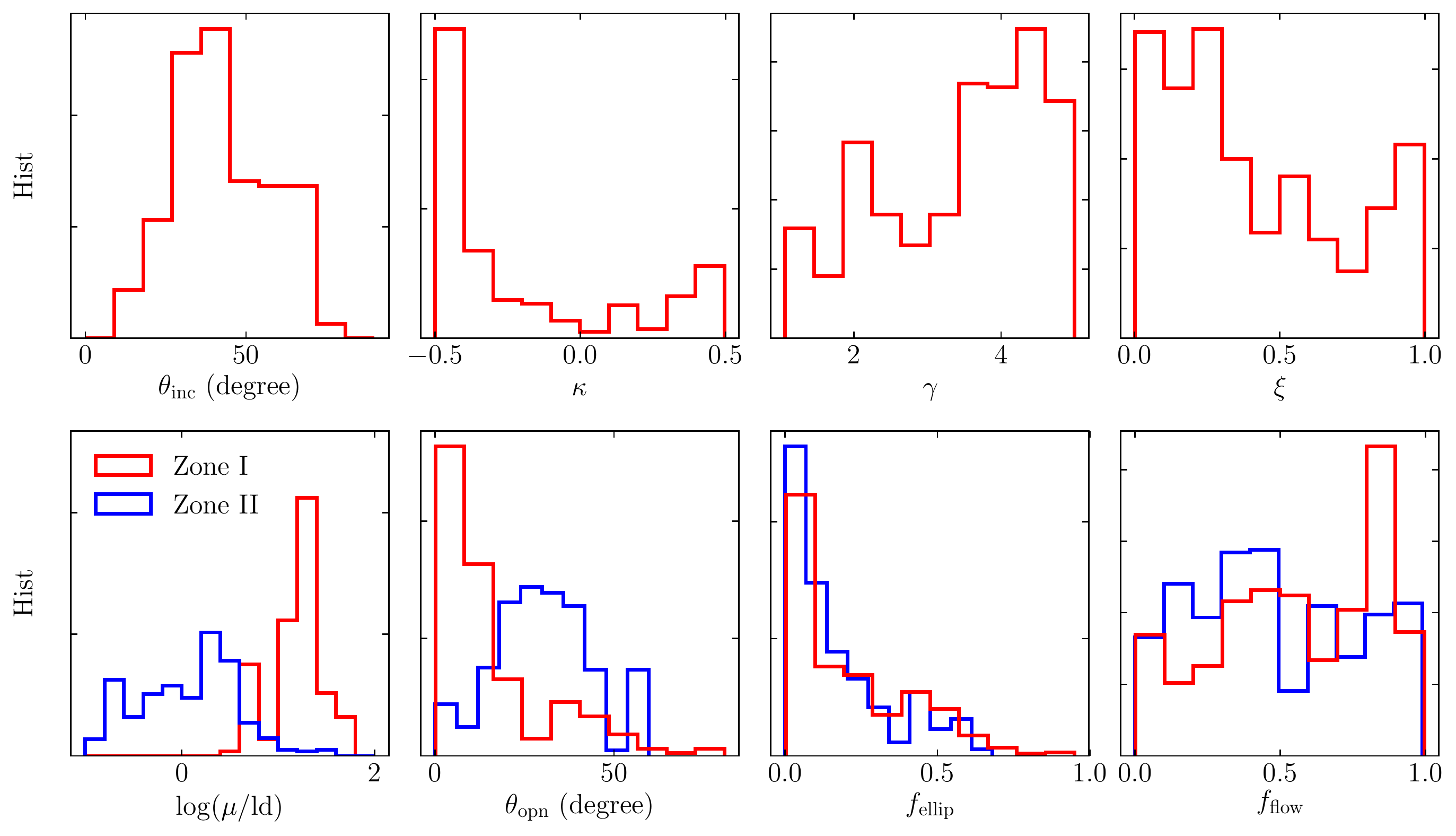}
\caption{Inferred posterior distributions of the selected main parameters for model $M3$. Top panels are for the 
common parameters of BLR zones I and II (see the schematic in Figure~\ref{fig_sch}). Bottom panels show the 
distributions of the parameters for zone I in red and zone II in blue. 
}
\label{fig_blrcmp}
\end{figure*}

An example of two-dimensional transfer function is shown in the top left panel of Figure~\ref{fig_ccf}.
It is slightly asymmetric with longer response at red side of the H$\beta$ profile. Such an asymmetric feature 
is more clearly seen in the velocity-binned delay map in the top right panel. This asymmetry is mainly caused by 
the anisotropic parameter $\kappa\neq0$ which means that particles' emissions depend on the locations 
and also by the dynamical parameter $f_{\rm ellip}<1$ which means that particles have 
inflow ($f_{\rm flow}<0.5$) or outflow ($f_{\rm flow}>0.5$) motion  (see Figure~\ref{fig_blrcmp}) . 
The bottom left panel of Figure~\ref{fig_ccf} plots the delay integral of transfer function $\Psi(v)$, in good agreement with 
the scaled mean H$\beta$ profile as expected from Equation (\ref{eqn_psi}).

\subsection{The Structure and Dynamics of the Two-zone BLR}
The best model $M3$ indicates that the BLR in Mrk~142 consists of two regions, consistent with the self-shadowing effects
of slim accretion disk models (\citealt{Wang2014b}). In Figure~\ref{fig_blrcmp}, we plot the inferred posterior distributions 
of several selected main parameters in model $M3$. The top panels shows the 
common parameters for both zones I and II (see the schematic in Figure~\ref{fig_sch}). The bottom panels show the 
distributions of the parameters for zone I in red and zone II in blue. The inclination angle is 
$41_{-11}^{+21}$. The anisotropic parameter $\kappa$ peaks at either $-0.5$ or $0.5$, which means that the observer 
sees the majority of emissions from either near side or far side of the BLR. The parameter $\gamma$ has a broad distribution
over (1, 5), but tends to peak at $\gamma=5$, indicating that the particles tend to concentrate near the outer face of 
the BLR disks. The distribution of $\xi$ is also broad and peak around $\xi=0$, which corresponds to a completely obscured 
half of the BLR below the equatorial plane.

From the bottom panels of Figure~\ref{fig_blrcmp}, we see that the distributions of the  dynamical parameters $f_{\rm ellip}$ and $f_{\rm flow}$
are roughly similar for zones I and II. However, the mean radius $\mu$ and the opening angle $\theta_{\rm opn}$ 
are different. Remarkably, the mean radius of zone I is clearly larger than that of zone II, in 
agreement with the theoretical model proposed by \cite{Wang2014b}. Because of the self-shadowing effects 
of the inner funnel of geometrically thick slim disks, the ionizing continuum flux received by zone II 
is significantly lower than that received by zone I, leading to a shrunken BLR size of zone II.
The ratio of the size scales of the two zones can be approximated by (\citealt{Wang2014b})
\begin{equation}
 \frac{\langle R \rangle_{\rm I}}{\langle R \rangle_{\rm II}}\approx 2\times \left(\frac{\mathscr{\dot M}}{50}\right)^{0.3}.
\end{equation}
{\cb Using the accretion rate $\log\mathscr{\dot M}=2.4_{-0.6}^{+1.3}$ for Mrk 142 obtained below, 
the anticipated ratio is $\log\left(\langle R \rangle_{\rm I}/\langle R \rangle_{\rm II}\right) = 0.5_{-0.2}^{+0.4}$. Our results give $\log\left(\langle R \rangle_{\rm I}/{\rm ltd}\right)=1.22_{-0.39}^{-0.21}$ and $\log\left(\langle R \rangle_{\rm II}/{\rm ltd}\right)=0.09_{-0.66}^{+0.44}$,}
marginally consistent with the anticipated value to within uncertainties.

\subsection{Comparison with the Cross-Correlation Analysis}
In Figure~\ref{fig_ccf}, we compare the velocity-binned time lags from our model fitting 
with these from CCF analysis over velocity bins chosen to be the same as those in \citetalias{Du2016b}. 
We only use the velocity bins with the maximum correlation coefficients $r_{\rm max}\geqslant0.7$.
At each velocity bin, the cross-correlation is calculated using the standard interpolated CCF method 
(\citealt{Gaskell1987}). The time delay is determined by either measuring the location $\tau_{\rm peak}$ of the 
CCF peak ($r_{\rm max}$) or the centroid $\tau_{\rm cent}$ of the points around the peak above the threshold $r \geqslant 0.8r_{\rm max}$.
As for model fitting, we calculate $\tau_{\rm peak}$ and $\tau_{\rm cent}$ by cross-correlating the observed 
continuum light curve with the reconstructed H$\beta$ light curves interpolated to the observed epochs.
There is a tendency that the time lags from model fitting are slightly shorter than these from CCF analysis.
We ascribe such an discrepancy to the reason that the CCF (between the observed continuum and H$\beta$ light curves) 
are broad and possibly multimodal, as seen from the bottom right panel of Figure~\ref{fig_ccf}. Nevertheless, 
the time lags from two approaches are consistent within uncertainties (at 2$\sigma$ confidence level).
The velocity-binned time lags show an asymmetric pattern with slightly longer lags at red side, which is usually regarded to 
be a signature of outflowing BLR (see the discussion in \citetalias{Du2016b}). Our modeling results in 
Figure~\ref{fig_blrcmp} show that the values of the dynamical parameter $f_{\rm flow}$ for both zones I and II distributes
over a broad range, meaning that both inflow and outflow can fit the RM data.

\begin{deluxetable}{lccl}
\tabletypesize{\footnotesize }
\tablecolumns{3}
\tablewidth{0.48\textwidth}
\tablecaption{\cb A summary for $f$ factor measurements.\label{tab_factor}}
\tablehead{
  \colhead{~~~~~$f$ factor~~~~~} &
  \colhead{~~~~~~~~~Value~~~~~~~~~} &
  \colhead{~~~Ref~~~}   &
  \colhead{~~~Note~~~}
}
\startdata
$\log f_{\rm RMS, \sigma}$  &  $\phm{-}0.74_{-0.16}^{+0.12}$   &  1      & \nodata\\
                            &  $\phm{-}0.45_{-0.09}^{+0.10}$   &  2     & \nodata\\
                            &  $\phm{-}0.71_{-0.11}^{+0.11}$   &  3   & \nodata \\
                            &  $\phm{-}0.77_{-0.13}^{+0.13}$   &  4        & \nodata \\
                            &  $\phm{-}0.63_{-0.12}^{+0.09}$   &  5     & \nodata \\
                            &  $\phm{-}0.80_{-0.12}^{+0.09}$   &  6         & classical bulges\\
                            &  $\phm{-}0.51_{-0.11}^{+0.09}$   &  6         & pseudo bulges\\ 
                            &  $\phm{-}0.57_{-0.07}^{+0.07}$   &  7         & dynamical modeling \\
                            &  $-0.06_{-0.52}^{+0.30}$         &  8         & dynamical modeling \\\hline
$\log f_{\rm RMS, FWHM}$    &  $\phm{-}0.18_{-0.13}^{+0.10}$   &  6         & classical bulges\\
                            &  $-0.15_{-0.15}^{+0.11}$         &  6         & pseudo bulges\\
                            &  $-0.40_{-0.56}^{+0.34}$         &  8         & dynamical modeling \\\hline
$\log f_{\rm mean, \sigma}$ &  $\phm{-}0.75_{-0.11}^{+0.09}$   &  6         & classical bulges\\
                            &  $\phm{-}0.28_{-0.20}^{+0.14}$   &  6         & pseudo bulges\\ 
                            &  $\phm{-}0.43_{-0.09}^{+0.09}$   &  7         & dynamical modeling \\                            
                            &  $\phm{-}0.07_{-0.52}^{+0.31}$   &  8         & dynamical modeling \\  \hline                          
$\log f_{\rm mean, FWHM}$   &  $\phm{-}0.11_{-0.16}^{+0.12}$   &  6        & classical bulges\\
                            &  $-0.30_{-0.22}^{+0.15}$         &  6        & pseudo bulges\\
                            &  $\phm{-}0.00_{-0.14}^{+0.14}$   &  7         & dynamical modeling \\                            
                            &  $-0.36_{-0.54}^{+0.33}$         &  8         & dynamical modeling \\                            
\enddata
\tablerefs{(1) \cite{Onken2004}, (2) \cite{Graham2011}, (3) \cite{Park2012}, (4) \cite{Woo2013}, 
           (5) \cite{Grier2013a}, (6) \cite{Ho2014}, (7) \cite{Williams2018}, and (8) this work.}
\end{deluxetable}

\subsection{The BH Mass in Mrk 142}
Mrk 142 was previously monitored by the LAMP project (\citealt{Bentz2009}).
The obtained H$\beta$ centroid lag is as short as $2.74_{-0.83}^{+0.73}$ days, in contrast to the anticipated lag of 
$\sim20$ days\footnote{However, \cite{Li2013} reanalyzed the same data (velocity-unresolved) with an approach similar to
this work and obtained an H$\beta$ lag of $\sim15$ days, consistent with the anticipated value. }, 
making it a significant outlier in the BLR size-luminosity relation 
(\citealt{Bentz2013}). {\cb The estimated BH mass is $\log\,(M_\bullet/M_\odot)=6.23_{-0.21}^{+0.13}$ using 
the H$\beta$ line dispersion of $859\pm102$ km~s$^{-1}$
from the RMS spectrum and a virial factor of $f_{\rm RMS, \sigma}=5.5$ (\citealt{Bentz2009}).}
In our new observations for Mrk 142, the measured H$\beta$ centroid lag is $7.9_{-1.1}^{+1.2}$ days 
through CCF analysis (\citetalias{Hu2015}) and the estimated BH mass
is $\log\,(M_\bullet/M_\odot)=6.59_{-0.07}^{+0.07}$ using the H$\beta$ FWHM from the mean spectrum and a virial 
factor of $f_{\rm mean, FWHM}=1$. 

{\cb Our model $M3$
yields a BH mass of $\log\,(M_\bullet/M_\odot)=6.23_{-0.45}^{+0.26}$ for Mrk 142, 
in remarkable agreement with \cite{Bentz2009}'s measurement. The resulting virial factor is
$\log f_{\rm mean, FWHM} = -0.36_{-0.54}^{+0.33}$ and $\log f_{\rm RMS, FWHM} =-0.40_{-0.56}^{+0.34}$ for the H$\beta$ FWHM measured 
from the mean and RMS spectra, respectively, and $\log f_{\rm mean, \sigma} = 0.07_{-0.52}^{+0.31}$ and $\log f_{\rm RMS, \sigma} =-0.06_{-0.52}^{+0.30}$
for the H$\beta$ line dispersion measured from the mean and RMS spectra, respectively. There have been a number of $f$ 
calibrations reported in the literature, mainly based on  H$\beta$ line dispersion measured from RMS spectra 
(e.g., \citealt{Onken2004, Graham2011, Park2012, Woo2013, Grier2013a}). The calibrated value ranges 
from $\log f_{\rm RMS, \sigma} = 0.45_{-0.09}^{+0.10}$ (\citealt{Graham2011}) to 
$\log f_{\rm RMS, \sigma} = 0.77_{-0.13}^{+0.13}$ (\citealt{Woo2013}).
\cite{Ho2014} calibrated $f$ factors for AGNs with classical and pseudo bulges separately based on
the notion that classical and pseudo bulges obey different $M_\bullet-\sigma_\star$ relations (\citealt{Kormendy2013}).
They presented $f$ factors for classical and pseudo bulges separately in cases of 
four widely used measures of H$\beta$ line widths, namely, FWHM and line dispersion 
from mean and RMS spectra. Table~\ref{tab_factor} summarizes the measurements of $f$ factor in the literature.
Despite the large uncertainties ($\sim0.4$ dex), our obtained $f$ factors tend to coincide with 
the factors for pseudo bulges calibrated by \cite{Ho2014}. }
The host galaxy of 
Mrk~142 is a late-type spiral galaxy and shows a strong bar in the nucleus (\citealt{Ohta2007}). 
Also, the surface brightness decomposition of the {\it HST} image does not detect a notable
bulge component (\citealt{Bentz2013}, \citetalias{Du2014}), probably implying that Mrk~142 may 
not host a classical bulge. 

{\cb Meanwhile, \cite{Pancoast2014b}, \cite{Grier2017}, and \cite{Williams2018} applied the BLR 
dynamical modeling analysis developed by \cite{Pancoast2014a} to a sample of AGNs. 
By combining the $f$ factor measurements obtained in the three studies, \cite{Williams2018}
reported the mean $f$ factors: $\log f_{\rm mean, FWHM}=0.00\pm0.14$, 
$\log f_{\rm mean, \sigma}=0.43\pm0.09$, and $\log f_{\rm RMS, \sigma}=0.57\pm0.09$ (see Table~\ref{tab_factor}).
Our measured $f$ factors for Mrk 142 are marginally consistent with these results to within uncertainties.
}

{\cb We estimate the dimensionless accretion rate according to the equation (\citetalias{Du2014}) 
\begin{equation}
\mathscr{\dot M} = 20.1\times\left(\frac{L_{\rm 5100}}{10^{44}\cos\theta_{\rm inc}}\right)^{3/2}\left(\frac{M_\bullet}{10^7M_\odot}\right)^{-2}.
\end{equation}
Using the 5100~{\AA} luminosity $\log L_{5100}=43.56\pm0.06$ (\citetalias{Du2015}), the inclination angle $\theta_{\rm inc}=41_{-11}^{+21}$ degrees,
and the BH mass $\log(M_\bullet/M_\odot)=6.23_{-0.45}^{+0.23}$, we obtain $\log \mathscr{\dot M}= 2.4_{-0.6}^{+1.3}$.}
This confirms the BH in Mrk 142  to be an SEAMBH accreting at a super-Eddington rate.

\section{Discussions}

\subsection{\cb The Continuum Reconstruction by the DRW Process}
The continuum light curve is reconstructed using the DRW process, which is found to be sufficiently adequate 
for large samples of AGN light curves on timescales of weeks to years (e.g., \citealt{Kelly2009, MacLeod2010, Zu2011, Zu2013, 
Andrae2013, Kozlowski2016a}). However, on short timescales of days, there is evidence for deviations from 
DRW process for high-cadence AGN light curves monitored by the {\it Kepler} telescope 
(\citealt{Mushotzky2011, Kasliwal2015, Kozlowski2016b}). \cite{Kelly2014} proposed to use the generic
continuous-time autoregressive moving average (CARMA) models to characterize the variability features of
a broad range of stochastic light curves. The DRW process is a special case of CARMA processes with autoregressive order $p=0$
and moving average order $q=0$. Using the package {\texttt{CARMAPACK}\footnote{Accessible at https://github.com/brandonckelly/carma\_pack.}} 
developed by \cite{Kelly2014}, we can choose the best order of $p$ and $q$ for CARMA processes by minimizing 
the AIC. We find that DRW process is still the favorable model compared with 
high-order of CARMA processes for the data of Mrk 142. In addition,  
using the Bayesian framework proposed by \cite{Li2018}, we perform comparison between the DRW model and the power spectral 
density (PSD) model with a single power-law. Note that the DRW model has a PSD $\propto 1/[1+(f/f_0)^2]$, where $f$ is 
the frequency and $f_0 = 1/2\pi\tau_{\rm d}$. We confirm that the DRW model is slightly preferable.

{\cb On the other hand, different continuum models mainly affect the short time-scale variability of the reconstructed 
continuum between measurement points (e.g., see \citealt{Li2018}). Such an effect will finally influence the amplitudes of the inferred 
parameter uncertainties. However, the inherent convolution operation in RM analysis (see Equation~\ref{eqn_rm})
will largely smooth the short time-scale variations. We therefore expect that the estimated uncertainties should not be 
significantly affected by the details of the chosen continuum models (see also discussions in \citealt{Skielboe2015} and \citealt{Fausnaugh2018}).}

\subsection{Anisotropic Emission of the Central Ionizing Source}
We only take into account the possibility that the anisotropic ionizing emission from the geometrically 
thick funnel in the inner region of slim accretion disks produce two-zone BLRs. Indeed, 
there are two additional anisotropic effects for accretion disks. First, the ionizing emissions 
strongly depend on the angle between the symmetric axis of disks and the direction toward BLR particles.
Second, the sizes of the ionizing source may be no longer negligible when in particular using 5100~{\AA} continuum as 
a surrogate for ionizing continuum {\cb (see also Section~\ref{sec_point})}. The first effect will cause the BLR to be thicker to compensate
the $\cos\theta$ dependence of the disk emission. To include the second effect, one needs to 
solve the structure of accretion disks and obtain the radial distribution of emissions. This will 
make the present model more complicated and the MCMC sampling more inefficient. We are thus content with the present 
simple treatments on the ionizing sources and defer the inclusion of these two effects to a separate paper. 

{\cb 
\subsection{Point-like Geometry of the Central Ionizing Source}
\label{sec_point}
We implicitly assume that the emission region of the 5100~{\AA} continuum is point-like.
However, multiwavelength RM observations on a handful of AGNs indeed detected 
time lags of the optical continuum variations with respect to the X-ray/UV variations (e.g., \citealt{Edelson2015, Edelson2017, Cackett2018,
Fausnaugh2018}).
This indicates that the emission region at 5100~{\AA} could be spatially extended.
We can estimate the characteristic radius for emission of the 5100~{\AA} continuum using the standard accretion disk
model. The local effective temperature of the accretion disk is written (e.g., \citealt{Laor2011})
\begin{equation}
T(r)=f(r, a)\left(\frac{3c^6}{8\pi G^2\sigma}\right)^{1/4}
     \frac{\dot M^{1/4}}{M_\bullet^{1/2}}r^{-3/4},
\end{equation}
where $r=R/R_{\rm g}$, $R_{\rm g}$ is the gravitational radius, $a$ is the BH spin, $\dot M=\mathscr{\dot M}L_{\rm Edd}/c^2$
is the mass accretion rate, $M_\bullet$ is the BH mass, $\sigma$ is the Stefan-Boltzmann constant, and 
$f(r, a)$ is a dimensionless factor on the order of unity that is set by the 
inner boundary condition and the relativistic effects. Regardless 
of $f(r, a)$ and using $\log\mathscr{\dot M}=2.4_{-0.6}^{+1.3}$ and $\log(M_\bullet/M_\odot)=6.23_{-0.45}^{+0.23}$,
the corresponding radius for the 5100~{\AA} emission is $R_{\rm 5100}=0.14_{-0.11}^{+0.60}$~ld. 
Note that for a slim disk, the presence of prominent radial advection reduces the effective temperature (\citealt{Abramowicz1988, Wang1999}), 
making the above estimate conservative. Considering that the inferred disk size from multiwavelength reverberation mapping observations 
is about 3 times larger than that predicted from the standard disk model (e.g., \citealt{Edelson2015}),  
the 5100~{\AA} emission radius $R_{5100}$ would be comparable with the mean radius of zone I, but much smaller than  
the mean radius of zone II (see Table~\ref{tab_val}). Spatial extension of the 5100~{\AA} emission region 
may lead the obtained BH mass to be underestimated. It is worth a detailed study for the influences of 
spatially extended 5100~{\AA} emission region on BH mass measurement. 
As discussed in the preceding section, for the sake of simplicity, we keep the assumption of point-like geometry 
of the 5100~{\AA} emission region and defer the detailed study to a future paper.
}

\subsection{Model Dependence of the Results}
{\cb In the present BLR models, the prescriptions for BLR properties are 
purely phenomenological and adopted only for the sake of simplicity}. 
This raises an issue as to whether the inferred results depend on the adopted model.
To address this issue, we need: 1) independent measurements from alternative approaches, 
and 2) model selections to evaluate the most probable model for BLRs. A major challenge
for performing model selections is that the existing BLR models (see the summary in Table~1 of \citealt{Wang2012})
invoke complicated physical processes, impeding an efficient MCMC inference.
The results from the three BLR models indeed imply that the obtained BH masses appear to be slightly different, although
they reproduce the RM data with different degrees of success.
Recently, \cite{Czerny2017} developed a self-consistent BLR model based on the failed radiatively accelerated
dusty outflow model (\citealt{Czerny2011}), which only invokes the basic physical parameters, such as BH mass
and spin, and accretion rate. The model is purely analytic and therefore apt for MCMC realization.
A comparison of the inferred parameters from this model and the present dynamical modeling will shed light into 
the issue as to model dependence of the results.

\subsection{Parameter Degeneracy}
The significant degeneracy in present models is among the BH mass, the inclination angle and the 
opening angle. There are two reasons causing this degeneracy (\citealt{Grier2017}). The first reason is from the model itself, 
such as the strong correlation between BH mass and inclination angle or opening angle, 
ascribed to the adopted disk-like geometry for the BLR (\citealt{Collin2006, Li2013}). The other reason is 
from the constraints by observation data. Similar to \cite{Pancoast2014b} and \cite{Grier2017}, we also find a tight
correlation between inclination and opening angles. Moreover, the values of these two angles are approximately equal. 
As pointed out by \cite{Grier2017}, the interpretations 
for such behavior are two-fold: first, to generate single-peaked line profiles, the opening angle should be larger than
the inclination angle; secondly, as the opening angle increases, the generated profiles tend to be flat in 
the core (e.g., \citealt{Netzer2010}), apparently incompatible with the observed line shapes.
Therefore, the observations require the opening angle as small as possible while still large enough to produce
single-peaked line profiles. As a result, the opening angle approximately equals to the inclination angle.

\subsection{BLR Dynamical Modeling}
The present dynamical model does not include possible systematic errors for 
the model assumptions. This issue can be overcome by comparing mass measurements against 
these from the other independent techniques, such as stellar dynamics and gas dynamics widely
used in quiescent galaxies. Unfortunately, the objects with both RM monitoring and the other 
independent measurements are still extremely few (\citealt{Peterson2014}). On the other hand,
new techniques such as spectro-astrometry (\citealt{Gnerucci2010, Stern2015}) and  
spectro-interferometry (\citealt{Kraus2012, Petrov2012}) are in the process of development with the purpose of 
spatially resolving gas dynamics surrounding the central BHs using the current ground-based 10m class telescopes
(\citealt{Gnerucci2011, Gnerucci2013, Rakshit2015}). Hopefully, 
in the near future, there will be sufficient data sample with independent mass measurements
that allow us to explore the systematic errors of our dynamical models.

{\cb 
\subsection{Comparison with the Maximum Entropy Method}
The maximum entropy method (MEM, \citealt{Horne1994}) is also widely used to derive transfer functions of BLRs
and probe structure and dynamics of BLRs. \citetalias{Xiao2018} presented the transfer function for Mrk 142 by applying
MEM to the same RM data used in this paper. The obtained transfer function exhibits a 
major response around 5-10 days (see Figure 11 in \citetalias{Xiao2018}), seemingly distinct from the transfer function 
derived from our dynamical modeling analysis (see the top right panel of Figure~\ref{fig_ccf}). 
However, this is not the case because of the following reasons. First, MEM solves a {\it modified} equation
compared to Equation (\ref{eqn_rm}) (regardless of the no-linear response)
\begin{equation}
f_l(v, t) = \bar f_l(v) + \int \Psi_{\rm MEM}(v, \tau)\left[f_c(t-\tau)-\bar f_c\right]d\tau,
\end{equation}
where $\bar f_l(v)$ and $\bar f_c$ are considered to be the constant background terms (\citealt{Horne1994}).
MEM employs maximum entropy regularization to find the smoothest solutions $\Psi_{\rm MEM}(v, \tau)$, $\bar f_l(v)$, and $\bar f_c$
that best fit the observed data. In real implementation, the derived $\bar f_l(v)$ and $\bar f_c$
usually include contributions from the nonvariable part of the broad emission line and continuum, which
cannot be attributed to background contributions (\citealt{Wanders1995}). As a result, $\Psi_{\rm MEM}$ is sensitive to 
changes in the responses of the BLR but insensitive to the total responses of the BLR. In this sense, it is more appropriate to 
call $\Psi_{\rm MEM}$ ``marginal transfer function''.

Second, MEM uses a free parameter to control the trade-off between smoothness of the solutions and goodness of 
fitting to the data. In practice, the value of this free parameter is chosen by eye to achieve the best compromise.
Sharp features in transfer functions will generally be smeared out by MEM (see also discussions in \citealt{Pancoast2018}),
so it is not straightforward to perform a direct, quantitative comparison with the results from dynamical modeling. 
}

\section{Conclusions}
We employ the recently developed dynamical modeling for broad-line regions
to analyze the RM data of broad H$\beta$ line and 5100~{\AA} continuum for Mrk 142 monitored between 2012 and 2013. 
The BH mass is self-consistently measured without resort to the virial factor required 
in the traditional RM analysis through the cross-correlation method.  
The main results are as follows:
\begin{enumerate}
 \item We apply three BLR models to fit the RM data of Mrk~142 and find that the best model
 is a two-zone model (see the schematic in Figure~\ref{fig_sch}), consistent with the 
 theoretical BLR model proposed by \cite{Wang2014b}. The two zones may be caused by
 the anisotropic ionizing emission due to the self-shadowing of the slim accretion disk. 
 Interestingly, the obtained mean size of zone~I is larger than that of zone~II, also in
 agreement with the theoretical model. It is possible that a much more complicated one-zone BLR model 
 can also fit the data of Mrk 142. Still, our results are illustrative and application to 
 other SEAMBH objects is required to reinforce the scenario of the two-zone BLR model.
 
 \item The general geometry of H$\beta$ BLRs for Mrk~142 is described by 
 an inclined disk with an inclination angle of 
 $42_{-11}^{+21}$ degrees. The opening angles for zones I and II are $30_{-12}^{+14}$ degrees and 
 $10_{-5}^{+26}$ degrees, respectively, corresponding to a thick disk with a total height aspect of 
 $h/r\sim0.6$.

 \item {\cb The obtained BH mass is $\log (M_\bullet/M_\odot)=6.23_{-0.45}^{+0.26}$, resulting in
 a virial factor of $\log f_{\rm mean, FWHM} = -0.36_{-0.54}^{+0.33}$ and $\log f_{\rm RMS, FWHM} =-0.40_{-0.56}^{+0.34}$ for the H$\beta$ FWHM measured 
from the mean and RMS spectra, respectively, and $\log f_{\rm mean, \sigma} = 0.07_{-0.52}^{+0.31}$ and $\log f_{\rm RMS, \sigma} =-0.06_{-0.52}^{+0.30}$
for the H$\beta$ line dispersion measured from the mean and RMS spectra, respectively.
These values are consistent to within uncertainties with previous measurements by similarly applying dynamical modeling 
to a dozen of AGNs (\citealt{Pancoast2014b, Grier2017, Williams2018}).
 Our obtained factors appear to coincide with the calibrations by \cite{Ho2014}  using the $M_\bullet-\sigma_\star$ relation 
 for pseudo bulges. If taking into account the intrinsic scatter ($\sim$0.3 dex) of 
 the $M_\bullet-\sigma_\star$ relation (\citealt{Kormendy2013}), our obtained factors are 
 also marginally consistent with other calibrations that did not explicitly make a distinction between morphology of host bulges 
 (e.g., \citealt{Onken2004, Park2012, Woo2013, Grier2013a}; see Table~\ref{tab_factor}).
 The resulting dimensionless accretion rate is $\log\mathscr{\dot M}=2.4_{-0.6}^{+2.3}$,}
 confirming that the BH in Mrk~142 is an SEAMBH accreting at super-Eddington rate.

\end{enumerate}

We end by remarking that the present dynamical modeling for BLRs is still at an early stage of infancy. Nevertheless,
our application to Mrk 142 along with previous applications to a dozen of AGNs (\citealt{Pancoast2012, Brewer2011b, 
Pancoast2014b, Grier2017, Williams2018}) 
is enlightening. Compared with the traditional CCF approach, direct modeling of the BLR structure and dynamics 
can reveal much more information in the RM data and most importantly offers an approach for BH mass measurements 
without the need of invoking the virial factor. Future improvements of the dynamical modeling should address the issue of the 
associated systematic errors and incorporate physical processes (such as photoionization and radiation pressure).

\acknowledgements{We thank the referee for useful suggestions that improve the manuscript. 
We acknowledge the support of the staff of the Lijiang 2.4 m telescope. 
Y.R.L thanks Anna Pancoast and Brendon Brewer for useful discussions on BLR dynamical modeling.
This research is supported in part by the National Key R\&D Program of China (2016YFA0400700), 
by the CAS Key Research Program (KJZDEW-M06), and 
by grant No. NSFC-1113006, and -U1431228 from the National Natural Science Foundation of China.
Y.R.L. acknowledges financial support from the National Natural Science Foundation of China through grant No. 11570326 
and from the Strategic Priority Research Program of the Chinese Academy of Sciences grant No. XDB23000000. 
L.K.X.  acknowledges financial support from the Light of West China Program (Y7XB016001) and from
the National Natural Science Foundation of China through grant No. 11703077.
L.C.H. acknowledges financial support from Peking  University, the  Kavli  Foundation, and from
the National Natural Science Foundation of China through grant No. 11473002 and also No. 11721303.
B.W.H acknowledges financial support from the National Key R\&D Program of China (2017YFA0402703).
All the calculations in this work used the computing clusters at the Computer Center of 
the Institute of High Energy Physics.}

\software{\cb BRAINS \url{https://github.com/LiyrAstroph/BRAINS}}

\begin{figure*}
\centering 
\includegraphics[width=0.9\textwidth]{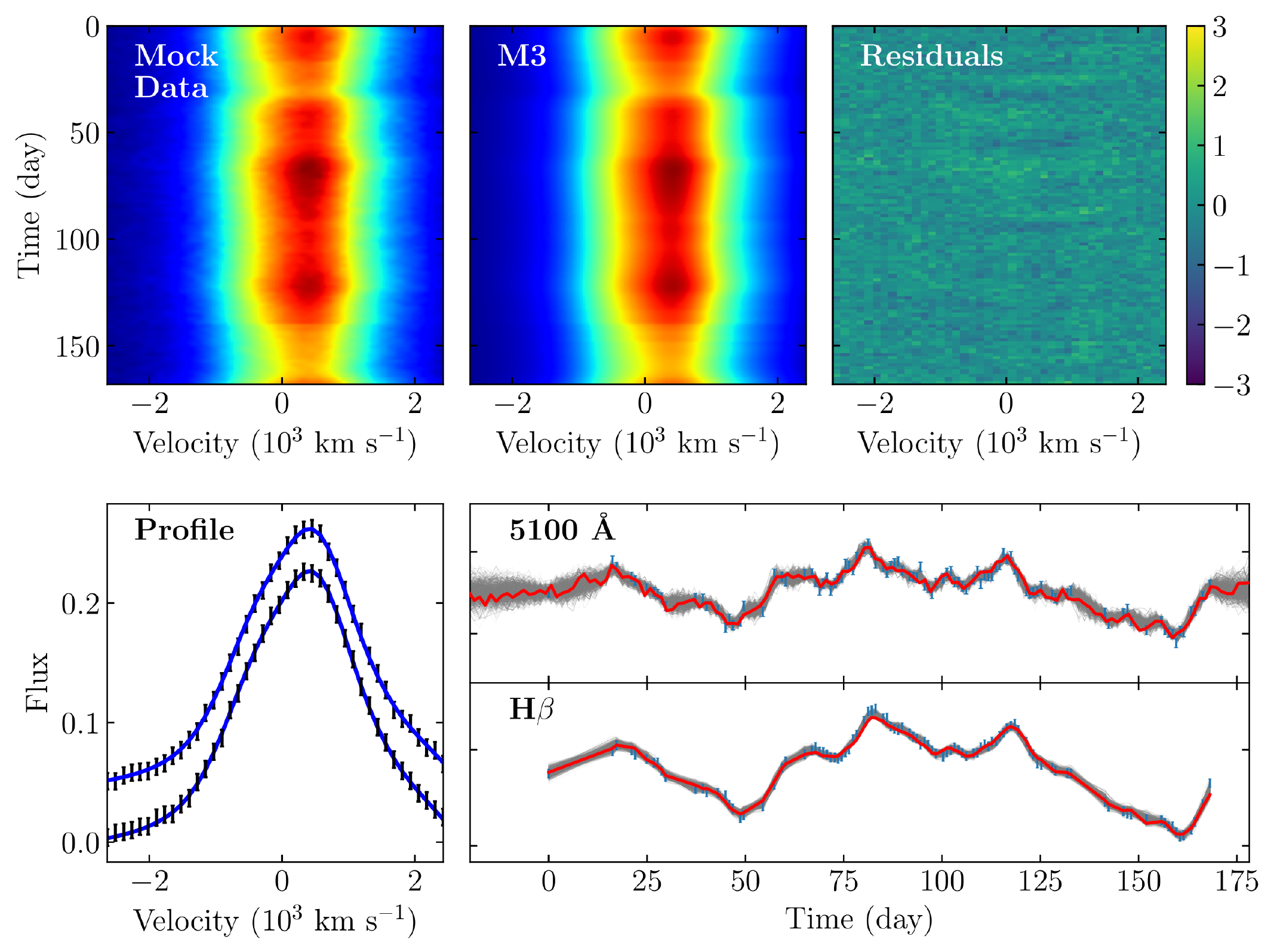}
\caption{The same as Figure~\ref{fig_m1}, but for fits to mock data generated using BLR model $M3$. }
\label{fig_sim}
\end{figure*}

\begin{figure*}
\centering 
\includegraphics[width=0.9\textwidth]{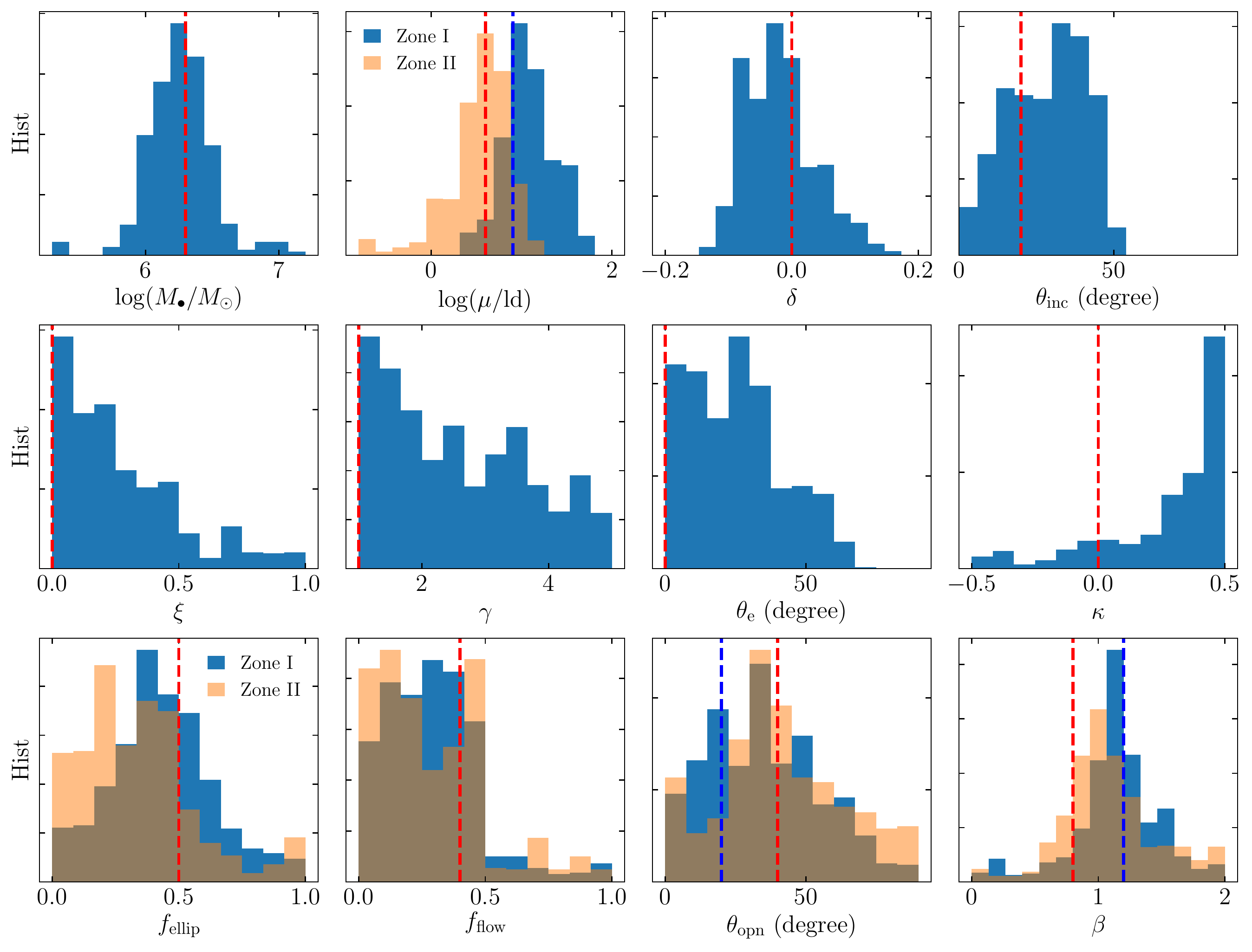}
\caption{Posterior distributions of the major parameters of model $M3$ inferred from the mock data shown in Figure~\ref{fig_sim}. Vertical dashed lines 
represent the input values for zone I (in blue) AMD zone II (in red). For the common parameters of zones I and II, red dashed line 
represent the input values. The parameters $f_{\rm ellip}$ and $f_{\rm flow}$ for zones I and II have the same input values.}
\label{fig_hist_sim}
\end{figure*}

{\cb
\appendix 

\section{Inferred Parameter Values for the Three BLR Models}
In Table~\ref{tab_val}, we summarize the inferred values of the major 
parameters for all three models. The best estimates are taken to be 
the median values of the posterior distributions and the uncertainties are 
taken from the 68.3\% confidence intervals.

\section{A Validity Test of the Code \texttt{BRAINS}}
To test the validity of our code, we generate mock data with the same cadence and spectral resolution as the RM data of 
Mrk 142  using model $M3$ (the two-zone 
BLR model). Figure \ref{fig_sim} shows the fitting results, and Figure \ref{fig_hist_sim} 
shows comparison between the posterior distributions of major parameters of $M3$ and 
the input values. As can be seen, the posterior distributions are generally consistent with 
the input values.
}

%
\begin{deluxetable*}{lccc}
\tabletypesize{\footnotesize }
\tablecolumns{4}
\tablewidth{1.0\textwidth}
\tablecaption{Inferred parameters for models $M1$, $M2$, and $M3$.\label{tab_val}}
\tablehead{
  \colhead{Parameter}
& \colhead{~~~~~~~~~~~~~~$M1$~~~~~~~~~~~~~~}
& \colhead{~~~~~~~~~~~~~~$M2$~~~~~~~~~~~~~~}
& \colhead{~~~~~~~~~~~~~~$M3$~~~~~~~~~~~~~~}
}
\startdata
$\delta$                       & $-0.17_{-0.10}^{+0.10}$        & $-0.17_{-0.13}^{+0.15}$       & $-0.11_{-0.08}^{+0.13}$    \\
$\log(\mu/\rm{ld})$            & $\phm{-} 0.92_{-0.27}^{+0.31}$ & \nodata                       & $0.09_{-0.66}^{+0.44}$   \\
$\beta$                        & $1.7_{-0.1}^{+0.1}$            & \nodata                       & $1.5_{-0.5}^{+0.3}$          \\
$F$                            & $0.09_{-0.02}^{+0.02}$         & \nodata                       & $0.24_{-0.21}^{+0.25}$           \\
$\log(\mu_{\rm I}/\rm ld)$     & \nodata                        & \nodata                       & $1.22_{-0.39}^{+0.21}$     \\
$\beta_{\rm I}$                & \nodata                        & \nodata                       & $1.1_{-0.2}^{+0.3}$            \\
$F_{\rm I}$                    & \nodata                        & \nodata                       & $0.15_{-0.05}^{+0.09}$            \\
$\rho_{\rm I}$                 & \nodata                        & \nodata                       & $0.38_{-0.10}^{+0.10}$           \\
$\alpha$                       & \nodata                        & $1.2_{-0.1}^{+0.1}$           & \nodata              \\
$\log(R_0/\rm ld)$             & \nodata                        & $0.5_{-0.4}^{+0.2}$           & \nodata            \\
$F_{\rm in}$                   & \nodata                        & $0.10_{-0.06}^{+0.11}$        & \nodata             \\
$\log F_{\rm out}$             & \nodata                        & $0.99_{-0.01}^{+0.01}$        & \nodata            \\
$\theta_{\rm inc}$ (degree)    & $45_{-12}^{+10}$               & $37_{-19}^{+10}$              & $41_{-11}^{+21}$            \\
$\theta_{\rm opn}$ (degree)    & $36_{-11}^{+10}$               & $43_{-10}^{+11}$              & $30_{-12}^{+14}$             \\
$\theta_{\rm opn, I}$ (degree) & \nodata                        & \nodata                       & $10_{-5}^{+26}$            \\
$\kappa$                       & $-0.36_{-0.09}^{+0.08}$        & $-0.48_{-0.01}^{+0.04}$       & $-0.38_{-0.09}^{+0.73}$        \\
$\gamma$                       & $4.44_{-1.5}^{+0.4} $          & $4.3_{-2.8}^{+0.5}$           & $3.7_{-1.6}^{+0.8}$             \\
$\xi$                          & $0.06_{-0.05}^{+0.10}$         & $0.18_{-0.11}^{+0.09}$        & $0.33_{-0.25}^{+0.52}$           \\
$\log(M_\bullet/M_\odot)$      & $5.90_{-0.31}^{+0.31}$         & $6.34_{-0.25}^{+0.37}$        & $6.23_{-0.45}^{+0.26}$    \\
$f_{\rm ellip}$                & $0.02_{-0.01}^{+0.04}$         & $0.01_{-0.01}^{+0.02}$        & $0.10_{-0.07}^{+0.28}$             \\
$f_{\rm flow}$                 & $0.76_{-0.20}^{+0.16}$         & $0.69_{-0.28}^{+0.20}$        & $0.42_{-0.24}^{+0.44}$            \\
$f_{\rm ellip, I}$             & \nodata                        & \nodata                       & $0.14_{-0.12}^{+0.31}$            \\
$f_{\rm flow, I}$              & \nodata                        & \nodata                       & $0.56_{-0.33}^{+0.32}$            \\
$\theta_{e}$ (degree)          & $19_{-12}^{+13}$               & $5_{-4}^{+9}$                 & $27_{-22}^{+26}$           \\
$\sigma_{\rm turb}$            & $-2.2_{-0.5}^{+0.7}$           & $-2.1_{-0.5}^{+0.6}$          & $-2.0_{-0.7}^{+0.6}$            
\enddata
\end{deluxetable*}

\end{document}